\documentclass{aa}  

\usepackage[dvipsnames]{xcolor}
\usepackage{graphicx}
\usepackage{textcomp} 
\usepackage{txfonts}

\newcommand{\J}{J22564-5910}

\newcommand{\Msol}{$M_\odot$}
\newcommand{\Rsol}{$R_\odot$}
\newcommand{\Mdot}{$\dot{M}$}

\begin{document} 

\title{Looking into the cradle of the grave:\\ J22564-5910, a young post-merger hot subdwarf?}
\titlerunning{J22564-5910: a young post-merger hot subdwarf.}



   \author{Joris Vos\inst{1,2}
          \and
          Ingrid Pelisoli\inst{1, 3}
          \and
          Jan Budaj\inst{4}
          \and
          Nicole Reindl\inst{1}
          \and
          Veronika Schaffenroth\inst{1}
          \and
          Alexey Bobrick \inst{5}
          \and
          Stephan Geier\inst{1}
          \and
          J.J. Hermes \inst{6}
          \and
          Peter Nemeth \inst{2,7}
          \and
          Roy \O{}stensen\inst{8}
          \and
          Joshua S. Reding \inst{9}
          \and
          Murat Uzundag \inst{10, 11}
          \and
          Maja Vu\v{c}kovi\'{c}\inst{10}
          }

    \institute{
    Institut f\"{u}r Physik und Astronomie, Universit\"{a}t Potsdam, Haus 28, Karl-Liebknecht-Str. 24/25, D-14476 Potsdam-Golm, Germany
    \and
    Astronomical Institute of the Czech Academy of Sciences, CZ-251\,65, Ond\v{r}ejov, Czech Republic
    \and
    Department of Physics, University of Warwick, Coventry, CV4 7AL, UK
    \and
    Astronomical Institute of the Slovak Academy of Sciences, 05960 Tatranska Lomnica, The Slovak Republic
    \and
    Lund University, Department of Astronomy and Theoretical physics, Box 43, SE 221-00 Lund, Sweden
    \and
    Department of Astronomy, Boston University, 725 Commonwealth Ave., Boston, MA 02215, USA
    \and 
    Astroserver.org, 8533 Malomsok, Hungary
    \and
    Department of Physics, Astronomy, and Materials Science, Missouri State University, Springfield, MO 65804, USA
    \and
    University of North Carolina at Chapel Hill, Department of Physics and Astronomy, Chapel Hill, NC 27599, USA
    \and
    Instituto de F\'{\i}sica y Astronom\'{\i}a, Universidad de Valpara\'{\i}so, Gran Breta\~{n}a 1111, Playa Ancha, Valpara\'{\i}so 2360102, Chile
    \and
    European Southern Observatory, Alonso de Cordova 3107, Santiago, Chile
    }

   \date{Received September 15, 1996; accepted March 16, 1997}

 
  \abstract
   {We present the discovery of \J, a new type of hot subdwarf (sdB) which shows evidence of gas present in the system and has shallow, multi-peaked hydrogen and helium lines which vary in shape over time. All observational evidence points towards \J\ being observed very shortly after the merger phase that formed it.}
   {Using high-resolution, high signal-to-noise spectroscopy, combined with multi-band photometry, Gaia astrometry, and TESS light curves, we aim to interpret these unusual spectral features.}
   {The photometry, spectra and light curves are all analysed, and their results are combined in order to support our interpretation of the observations: the likely presence of a magnetic field combined with gas features around the sdB. Based on the triple-peaked H lines, the magnetic field strength is estimated and, by using the {\sc shellspec} code, qualitative models of gas configurations are fitted to the observations.}
   {All observations can either be explained by a magnetic field which enables the formation of a centrifugal magnetosphere, or a non-magnetic hot subdwarf surrounded by a circumstellar gas disk/torus. Both scenarios are not mutually exclusive and both can be explained by a recent merger.}
   {\J\ is the first object of its kind. It is a rapidly spinning sdB with gas still present in the system. It is the first post-merger star observed this early after the merger event, and and as such is very valuable system to test merger theories.  If the magnetic field can be confirmed, it is not only the first magnetic sdB, but it hosts the strongest magnetic field ever found in a pre-white dwarf object. Thus, it could represent the long-sought for immediate ancestor of strongly magnetic WDs. }


   \maketitle
%

\section{Introduction}
Hot subdwarf-B (sdB) stars are core helium-burning stars with M $\simeq$ 0.5$M_{\odot}$ and hydrogen envelopes too thin to sustain hydrogen-shell burning (M$_{env}$ $<$ 0.01 $M_{\odot}$, \citealt{Heber2016}). They are of particular interest for binary evolution as they can only be formed through binary interaction mechanisms \citep{Pelisoli2020}. The three binary formation channels that are thought to contribute significantly to the population are \citep{Han2002, Han2003}:
(1) Common envelope (CE) ejection. In this case, the sdB star forms from the core of a red giant branch  (RGB) star which has lost its envelope due to a companion and ignited helium. If mass transfer on the RGB is unstable, the binary will enter a common envelope phase, and the orbit will shrink until the envelope is ejected, resulting in a short period sdB binary with a main sequence (MS) or white dwarf (WD) companion.
(2) Stable mass transfer. If mass transfer on the RGB in the previous scenario is stable, the sdB will lose its envelope during Roche-lobe overflow (RLOF), resulting in a wide sdB + MS binary\citep[e.g.][]{Vos2020}.
(3) A merger of two low-mass He-WDs or a He-WD with an M dwarf (dM), which will result in a single sdB star \citep{Webbink1984}.

Many studies have attempted to model this He-WD merger channel and produce the observed population of single sdBs and their hotter counterparts, the O-type subdwarf (sdO) stars \citep[e.g.][]{Iben1990, Saio2000, Zhang2012}. Two main problems remain in these models: (1) reproducing the atmospheric composition of the H-rich sdB stars, and (2) spinning down the merger products. Recent models manage to match the observed H, He and CNO composition of the observed single sdB stars \citep{Hall2016}. However, He WD merger models still cannot explain the observed rotational velocities \citep{Schwab2018}.

A suggested explanation for the discrepancy in rotational velocities between observed single sdBs and the models is the effect of magnetic fields. Magnetic coupling between the merger product and remaining mass around it could rapidly decrease the rotational velocity of the newborn sdB star \citep{Iben1986, Schwab2018}. Furthermore, the strong atmospheric composition anomalies found in hot subdwarfs have been linked to magnetic fields as they are similar to the anomalies found in magnetic main sequence Ap/Bp stars \citep{Landstreet2004}.

Magnetic fields are known to exist in hot stars without deep outer convective zones on the main sequence, typically explained by mergers \citep{Schneider2019} or being primordial \citep{Neiner2015}, and on the WD cooling track, usually explained by being primordial or related to CE evolution \citep{Tout2008, Ferrario2015}. In between, however, only weak magnetic fields are suggested in a few post-AGB stars \citep[e.g.][]{Sabin2015}, as well as in central binaries in planetary nebulae \citep[e.g.][]{Jordan2005}, and their existence is still debated \citep[e.g.][]{Jordan2012, Leone2014}. Recently \citet{Momany2020} discovered spots on extreme horizontal branch stars in globular clusters, potentially attributed to magnetic fields. However conclusive proof of magnetic fields in those objects is still lacking. Detection of magnetic fields in hot subdwarfs, which will evolve into hot white dwarfs, could be very helpful in understanding the global magnetic field of the host star as it changes due to stellar evolution \citep{Landstreet2012}.

Surveys aimed at detecting magnetic fields in hot subdwarfs have found several candidates \citep[e.g.][]{Elkin1996, OToole2005, Mathys2012, Heber2013}. Still, a careful reanalysis of the observations indicates that magnetic fields of kilogauss (kG) strength might be very rare or completely absent in hot subdwarfs \citep{Landstreet2012}. Currently, no magnetic fields have been conclusively detected in cool sdBs or horizontal branch stars \citep{Mathys2012}.

In this article, we present the discovery of \J\ (RA\,=\,22:56:24.30, Dec\,=\,--59:10:14.38). This system is an sdB star with very unusual spectral features. We show that the most probable interpretation for it is that \J\ is a young merger product, with an active magnetic field and gas present in the system.

\section{Spectral energy distribution}

\begin{table}
\caption{Photometry of \J\ collected from SKYMAPPER, GAIA, APASS, 2MASS and WISE}
\label{tb:photometry}
\centering
\begin{tabular}{lrrr}
\hline\hline
\noalign{\smallskip}
Band    &   Magnitude    &   Error   \\
        &  mag      &   mag \\\hline
\noalign{\smallskip}
SKYMAPPER $u$  &  14.343   &  0.012  \\
SKYMAPPER $v$  &  14.204   &  0.012  \\
SKYMAPPER $g$  &  14.153   &  0.008  \\
SKYMAPPER $r$  &  14.330   &  0.013  \\
SKYMAPPER $i$  &  14.617   &  0.005  \\
SKYMAPPER $z$  &  14.817   &  0.015  \\
GAIA3 $G$      &  14.2461  &  0.0031 \\
GAIA3 $BP$     &  14.2159  &  0.0049 \\
GAIA3 $RP$     &  14.2704  &  0.0053 \\
APASS $B$      &  14.281   &  0.029  \\
APASS $V$      &  14.261   &  0.022  \\
APASS $G$      &  14.193   &  0.021  \\
APASS $R$      &  14.452   &  0.052  \\
APASS $I$      &  14.678   &  0.054  \\
2MASS $J$      &  14.269   &  0.032  \\
2MASS $H$      &  14.290   &  0.053  \\
2MASS $KS$     &  14.082   &  0.064  \\
WISE $W1$      &  14.224   &  0.053  \\
WISE $W2$      &  14.177   &  0.015  \\
\hline
\end{tabular}
\end{table}

The photometric spectral energy distribution (SED) of \J\ can be used to estimate the effective temperature of the sdB and check for the possible close surrounding matter. Literature photometry from SKYMAPPER \citep{Wolf2018}, Gaia EDR3 \citep{GaiaEDR3, Riello2020}, APASS DR9 \citep{Henden2015}, 2MASS \citep{Skrutskie2006} and WISE W1 and W2 from the unWISE survey \citep{Schlafly2019} are used. There is also a Galex NUV measurement available, but this is not included in the fit for two main reasons. The GALEX UV photometry is not very reliable at the bright end, and the UV emission of sdB stars is very sensitive to metallicity \citep{Heber2016} and potential reddening from surrounding dust. All used photometry is shown in Table.\,\ref{tb:photometry}.

Using the Gaia parallax \citep{Lindegren2020}, the radius and luminosity of the sdB star can be constrained. For \J\ the distance obtained by inverting the parallax is $d = 646 \pm 13$ pc. The parallax zero point offset of \citep{Lindegren2020b} was applied before inverting the parallax. The Gaia RUWE factor is 1.032, which suggests a reliable astrometric solution, particular taking into account that variability also causes increase in RUWE \citep{Belokurov2020}. Furthermore, we check the reddening from the dust maps of \citet{Lallement2019}, which predict a reddening of E(B-V) = 0.015 $\pm$ 0.01 in the direction of \J. It has to be noted that these maps would not take the local dust in the system into account, and can thus not be used to constrain the SED fit.

\begin{figure}
    \includegraphics[width=\linewidth]{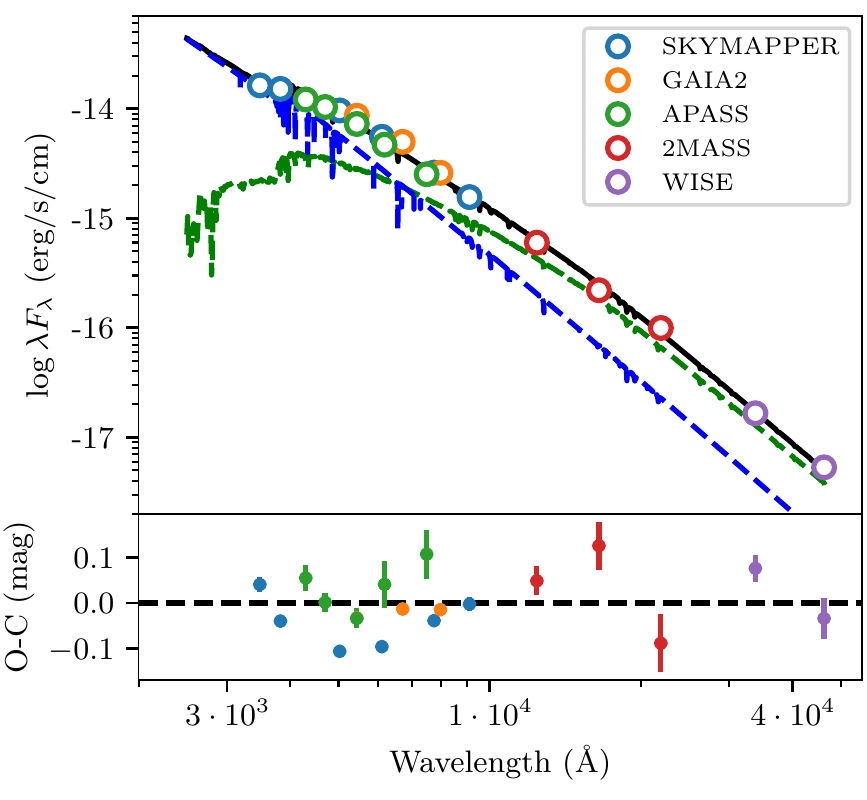}
    \caption{The photometric SED of \J\ obtained from literature photometry. The best-fitting model is shown in full black line, with the contribution of the sdB shown in blue dashed line. The contribution of the IR excess, likely from a disk, is shown in green dashed line. The bottom shows the O-C between the observations and the best fitting model.}\label{fig:sed_fit}
\end{figure}

To fit the SED of the sdB star, models from the T\"ubingen NLTE Model-Atmosphere package (\citealt{Werner2003}, TMAP) are used. It is clear from the SED that there is a significant contribution of a cooler component (See Fig.\,\ref{fig:sed_fit}). This is most likely a disk-like structure. However, the SED fitting package used here can only include spherical components. Therefore, the IR excess is modelled as a cool star using both Kurucz atmosphere models \citep{Kurucz1979} and a simple black body. A Markov chain Monte-Carlo approach is used to find the global minimum and determine the error on the fit parameters. The error on the distance is propagated throughout the fit. The code used is included in the {\sc speedyfit} python package\footnote{https://github.com/vosjo/speedyfit}. A more detailed explanation of the SED fitting approach can be found in \citep{Vos2012, Vos2013, Vos2017}.

\begin{table*}
\caption{Results of the SED fit. The top line gives the results of the fit performed with the TMAP models in combination of Kurucz models to model the circumstellar matter, while in the bottom line the circumstellar matter is models by a black body. In the table the parameters for the circumstellar matter are marked with `disk'.}
\label{tb:sed_results}
\centering
\begin{tabular}{lrrrrrr}
\hline\hline
\noalign{\smallskip}
Model    &   T$_{\rm eff}$ sdB    &   R sdB &  T$_{\rm eff}$ disk    & R disk & E(B-V) \\
         &   (K)  &   (R$_{\odot}$)   &   (K)  &   (R$_{\odot}$)  &   \\\hline
\noalign{\smallskip}
TMAP \& Kurucz     & 23000 $\pm$ 3000 & 0.12 $\pm$ 0.04 & 6000 $\pm$ 1800 & 0.30 $\pm$ 0.05 & 0.04 $\pm$ 0.03 \\
TMAP \& Black body & 21000 $\pm$ 3000 & 0.15 $\pm$ 0.04 & 5000 $\pm$ 1500 & 0.33 $\pm$ 0.05 & 0.05 $\pm$ 0.03 \\
\hline
\end{tabular}
\end{table*}

The best fitting binary SED models using a Kurucz and black body model for the companion as given in Table\,\ref{tb:sed_results}. Both models are indicative of a rather cool sdB star combined with a cool component with a temperature between 5000 - 6000 K and a radius around 0.3 R$_{\odot}$. The fitted reddening is higher than the value obtained from \citep{Lallement2019}, but has a large error. The model of the cool component is almost certainly nonphysical, and the IR excess is likely caused by a disk-like structure, not a star. The reason for this is that such a star would require an unlikely combination of a high surface temperature with a very small radius. Secondly, if this would be a dwarf star or even an ultra hot Jupiter \citep[e.g.][]{Lillobox2014}, its spectral features should be visible in the spectra, which is not the case. It is clear that a more detailed model taking the possible geometry of the companion into account is necessary to adequately fit the SED. 
However, the temperature and radius of the sdB star are likely reliable, as they are supported by the presence of several strong \ion{He}{i} lines, and the absence of \ion{He}{ii} lines in the spectra. 

\section{Spectral analysis}\label{sec:spectral_analysis}

The original EFOSC2 spectra were too low in resolution to show many features and only covered a small wavelength range. They did, however, show very broad H and He lines. If caused by stellar rotation alone, they would require a $v\sin{i} \sim 1100$ km/s, which surpasses the critical velocity of sdBs. This discovery led to follow-up observations on 29-12-2018 using the Goodman spectrograph \citep{2004SPIE.5492..331C} mounted on the 4.1-m Southern Astrophysical Research (SOAR) telescope on Cerro Pach\'{o}n in Chile. Using a 930-line grating with a 0.45\,\arcsec\ slit, the SOAR spectra have a better resolution and broader wavelength coverage but still had insufficient S/N. They did, however, show indication of line splitting for the hydrogen lines, and clear emission profiles for the H$_{\alpha}$ line, confirming the suspicion of gas or a magnetic field being present in the system/star. Based on these observations, follow-up observing proposals at both UVES and X-SHOOTER were approved, to study the unusual spectral features of \J.

In total 6 extra spectra were obtained, 3 UVES spectra and 3 X-SHOOTER spectra. Two UVES spectra were taken back to back, followed by a third one, 1 month later. The X-SHOOTER spectra have 1 week and 1 month in between them. These combinations allow us to check for variability on different time scales. The X-SHOOTER spectra were taken with a setup that favoured a higher S/N in the UVB and VIS in exchange for limited calibration of the NIR arm. Therefore the NIR spectra are of little use and are not included in our analysis. Details of these observations are given in Table\,\ref{tb:observations}.

    

\begin{table*}[t]
    \centering
    \caption{The observing date, exposure time, signal to noise and resolution and of the reduced spectra of the UVES and X-SHOOTER observations of \J. In the last two collums the derived radial velocities based on the wings of the hydrogen lines are given. }\label{tb:observations}
    \begin{tabular}{lllllll}
    \hline\hline
    \noalign{\smallskip}
    MJD & Instrument & Exp. time (s) & S/N & Resolution & RV (km/s) & err (km/s) \\
    \hline
    \noalign{\smallskip}
    58697.202063    &  UVES      & 2114 & 42   &  24000  &  33  &  41  \\
    58697.229497    &  UVES      & 2114 & 57   &  24000  &   8  &  50  \\
    58731.004125    &  UVES      & 2114 & 55   &  24000  &  25  &  44  \\
    58766.001310    &  X-SHOOTER & 1560 & 89   &  11000  &  79  &  28  \\ 
    58773.062886    &  X-SHOOTER & 1560 & 122  &  11000  &   7  &  24  \\
    58804.105317    &  X-SHOOTER & 1560 & 97   &  11000  & -14  &  21  \\
    \hline
    \end{tabular}
    
\end{table*}

In Fig.\,\ref{fig:xshooter_merged_normalized_UVB} and \ref{fig:xshooter_merged_normalized_VIS} the normalized spectrum created by summing the three X-SHOOTER spectra in both the UVB and VIS arm is shown. The spectrum shows several interesting features. Two Balmer lines, H$_{\beta}$ and H$_{\gamma}$, show a very clear triple absorption peak structure in their core. The same triple peak structure is visible in H$_{\delta}$ to H$_{\eta}$, but not as strong. The H$_{\alpha}$  line shows a very clear emission core that is stronger than the absorption part of the line. In the blue part of the spectrum, the Calcium K line has a triple absorption peak structure with a very sharp absorption peak at the centre of the line. The centre absorption peak is interstellar in origin. The interstellar Ca-H line is visible near the centre of the H$_{\eta}$ line. Furthermore, there are several \ion{He}{i} lines visible. The \ion{He}{i} $\lambda$ 4471 line shows the same triple peak structure visible in some of the hydrogen lines, but the centre of the line is shifted with respect to the rest wavelength by roughly 5 \AA. The \ion{He}{i} lines at 4921 and 5015 \AA\ show an emission core and also appear shifted with respect to the rest wavelength. The \ion{He}{i} line at 5875 \AA\ also has a strong emission core but is roughly centred at its rest wavelength. At the end of the UVB arm of the spectrum there are some bumps visible that could be the Mg I triplet at 5167, 5173 and 5184 \AA. However the quality of the spectrum is not sufficient to confirm this.
The two sharp absorption lines in the red part of the line are the Sodium doublet \ion{Na}{D} $\lambda$ 5890 and 5896 \AA. Further in the red part of the spectrum the \ion{O}{i} triplet at 7774 \AA\ shows core inversion similar to many of the He lines. The spectrum also shows some sharp lines in the red part, for example, at 6450-6520, 6960-7160, 7320-7400 and 7850-8100 \AA. These lines are terrestrial and not related to the system.

\subsection{Spectral trails}
\begin{figure*}
    \centering
    \includegraphics[width=6in]{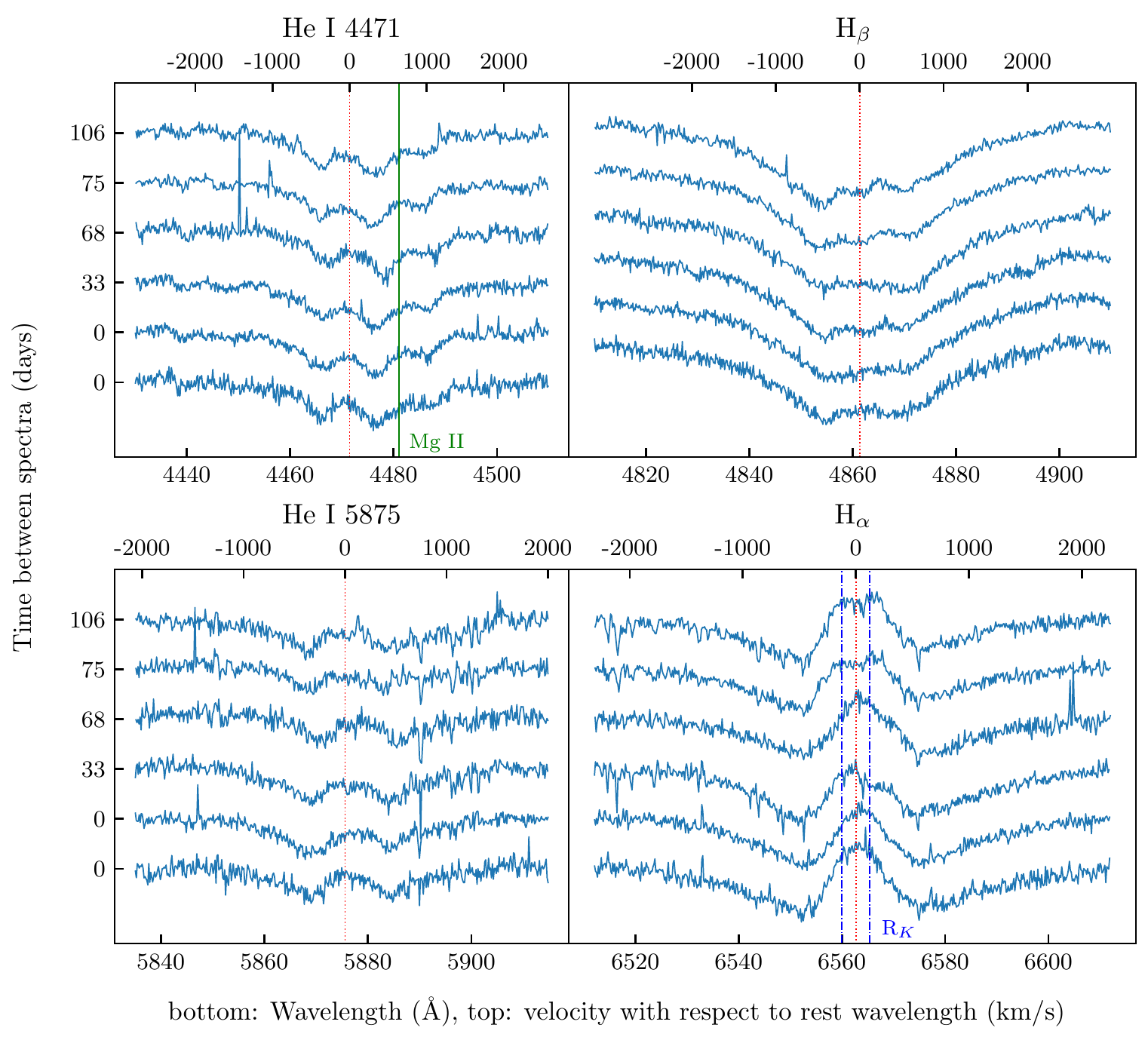}
    \caption{Spectral trail of hydrogen and helium lines visible in the spectrum of \J. The rest wavelength of each line is shown in red dotted line. On the top left plot the \ion{Mg}{ii} $\lambda$ 4483 line is marked in green. On the bottom right plot the location of R$_K$ is indicated in blue dash-dotted line (see Sect.\ref{s:magnetic_fields}). The spectra are shown in order of observations -- the time since the first spectrum is shown on the y axis.  The bottom three spectra were taken with UVES, while the top three spectra were taken with XSHOOTER. The first two UVES spectra were taken back to back. The axes on the bottom of the plot shows wavelength, while that on the top shows velocity compared to the rest wavelength. }\label{fig:spectral_trails_phot}
\end{figure*}

As multiple spectra are available, we can check if there is any change in the spectral features over time. In Fig.\,\ref{fig:spectral_trails_phot} the \ion{He}{i} lines at 4471 and 5875 \AA\ are shown together with H$_{\alpha}$, H$_{\beta}$, \ion{Ca}{k} and the \ion{O}{i} $\lambda$ 7774 line. The change in the H$_{\alpha}$ line is clearly visible. The emission core of the line varies between a single-peaked structure and a double-peaked structure. The \ion{He}{i} 5875 shows a similar but much weaker change; an emission peak that shifts from a double peak or flat-topped structure to a sharper single peak. The \ion{Ca}{k} doesn't show clear variations. But the the \ion{O}{i} $\lambda$ 7774 line does show variations, with the strongest absorption peak moves from blue to red shifted and back. Interesting about these latter two lines is that they are typically not visible in sdB spectra as they require lower temperatures. These could be produced in the circumstellar matter.

The \ion{He}{i} $\lambda$ 4471 line is a somewhat different feature. It can be interpreted as a triple absorption line that is shifted strongly from its central wavelength. Such a wavelength shift could be caused by the presence of a magnetic field (see Section\,\ref{s:magnetic_fields}). Another possible interpretation is that the line is a broad absorption line with an emission core centred on the rest wavelength of the line similar to the other two lines and that the rightmost absorption peak is caused by a different element (see Section\,\ref{s:disk}).

The actual periodicity of the line changes can not be determined from these spectra, but it is estimated to be on the order of several days to potentially even weeks.  Another limiting factor to this analysis is that, if the observed period is in fact due to rotation, the spectra should be affected by rotational smearing, given the long exposure times (up to a third of the period). This implies that we might not be sampling the spectral variability completely. An important notice is that the different lines vary with different periodicity. When comparing the H$_{\alpha}$ with the \ion{O}{i} $\lambda$ 7774 line, at time 0 both are central peaked. At time 33 days they have opposite absorption peaks with H$_{\alpha}$ blue shifted and \ion{O}{i} red shifted. At time 75 they are both red shifted. This would indicate that they have different origins or originate on different locations in the disk.


\begin{figure*}
    \centering
    \includegraphics[width=\linewidth]{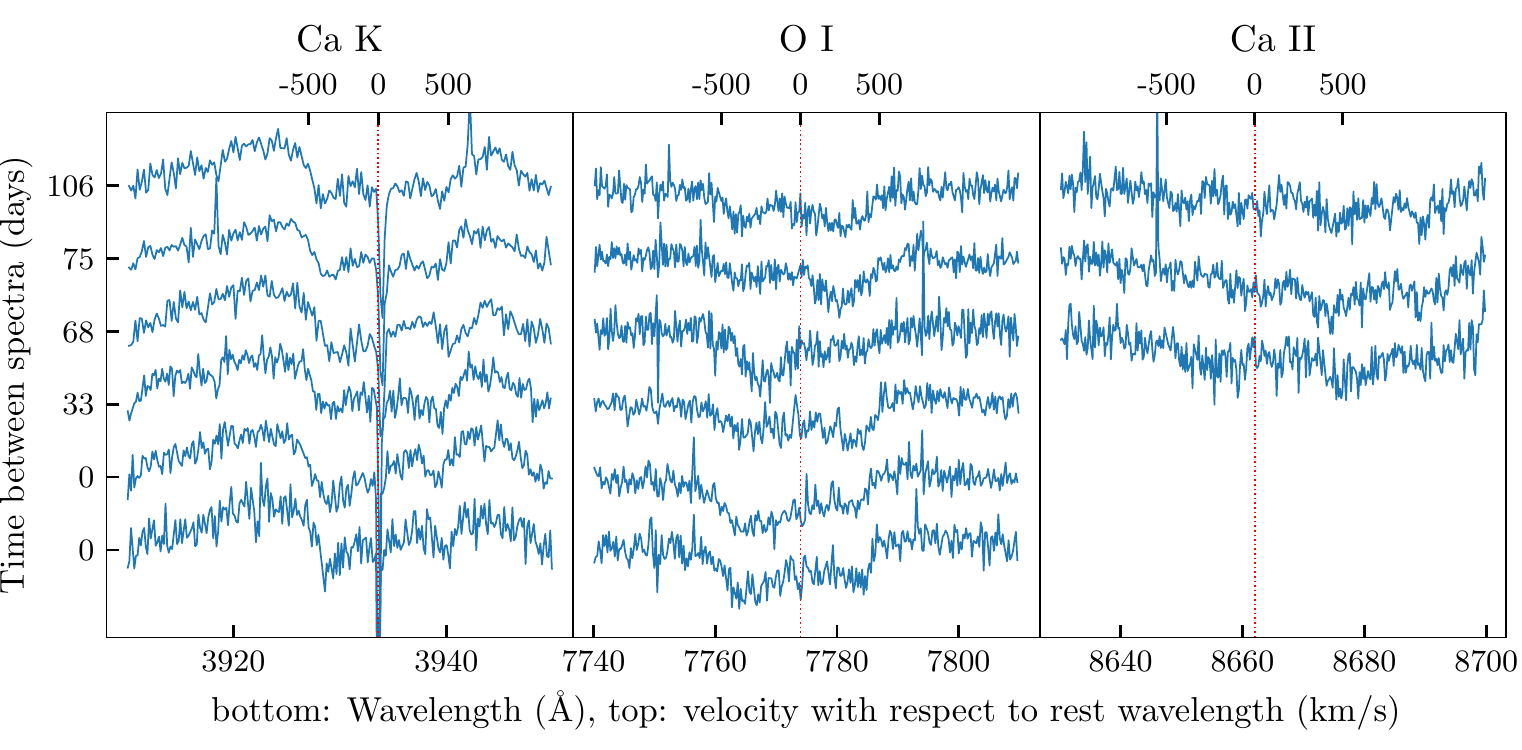}
    \caption{Same as Fig.\,\ref{fig:spectral_trails_phot}. Spectral trail of lines likely to originate in the circumstellar material visible in the spectrum of \J. For the \ion{Ca}{ii} line, the UVES spectra show only noise and are not shown here. The two other components of the Ca IR triplet are not visible in the spectra. }\label{fig:spectral_trails_non_phot}
\end{figure*}

\subsection{Radial velocity variations}\label{s:rv}
Given the spectra taken at different time intervals, it makes sense to attempt to check for radial velocity variations. However, this is complicated by the broad lines and varying line shapes. As the line cores of the hydrogen and helium lines vary strongly over time, they cannot be used to derive radial velocities. There are no clear, sharp lines visible in the spectrum belonging to the system, so the only remaining approach is to use the wings of the hydrogen lines. Different approaches were attempted, using cross-correlation with a template spectrum and the best-observed spectra, as well as fitting Gaussian functions to the wings. The most successful approach was Gaussian fitting, as it resulted in the least difference between radial velocities determined for different lines in the same spectrum.

To derive the radial velocities, hydrogen lines from H$_{\beta}$ to H$_{\eta}$ and H$_{10}$ have been used. For these lines, the line centres have been removed. The wavelength region that is excluded is determined by eye, by selecting the line part that varies the most in between the six spectra. Afterwards, a Gaussian is fitted to the wings of the same hydrogen line in all spectra, and the average value for its FWHM is used as a fixed value for the FWHM in the final fit. This way, all hydrogen lines are fitted, and the final radial velocity is the average of the RV of the different hydrogen lines. The error is calculated as the standard deviation between the different lines.

\begin{figure}
    \includegraphics[width=\linewidth]{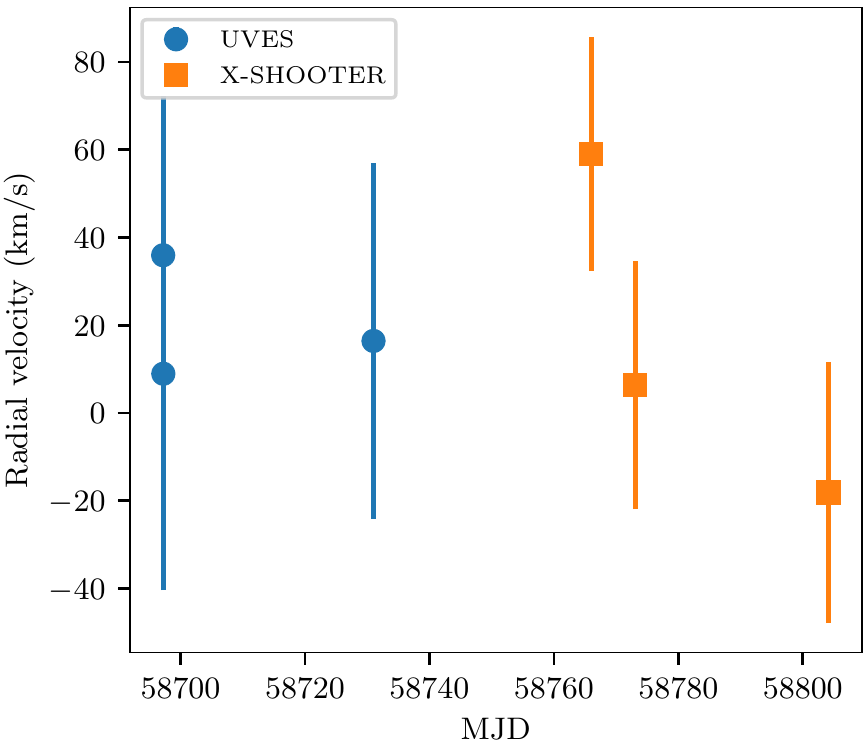}
    \caption{The radial velocities of \J\ as determined from the wings of the hydrogen lines. The first three measurements in blue are obtained from the UVES spectra, while the latter measurements are from the X-SHOOTER spectra. Apart from one measurement, all RVs are consistent with no variation. }\label{fig:radial_velocities}
\end{figure}

The resulting radial velocities are shown in Fig.\,\ref{fig:radial_velocities}, and are given in Table\,\ref{tb:observations}. As can be seen from that figure, almost all RVs are consistent with no significant RV variation. One spectrum, the first of the X-SHOOTER spectra, shows a possible deviation. However, from these observations, it is not possible to conclude whether the system is RV variable or not.

\section{TESS lightcurve}

\begin{figure}
    \includegraphics[width=\linewidth]{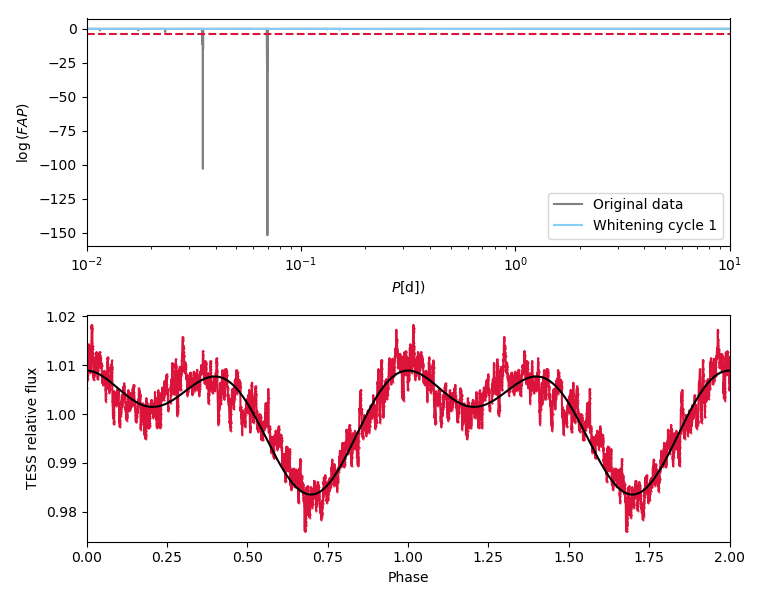}
    \caption{Top: Periodogram (grey) of the TESS light curve for \J. The light blue line is the re-calculated periodogram after the first whitening cycle. The red dashed line indicates $\log(FAP)=-4$. Bottom: Phase-folded (at the 0.069764\,d period, respectively) and averaged (every 50 points) TESS light curve.}\label{fig:tess}
\end{figure}

\J\ (TIC 220490049) was observed by the Transiting Exoplanet Survey Satellite \citep[TESS][]{ricker2015} during Sectors 01 and 28. Two-minute cadence data are not available, because the object was not included in the TESS target list, but full-frame images (FFI) available, with a 30-minute cadence for Sector 01 and 10-minute cadence for Sector 28. We downloaded a cutout of 50x50 pixels using TESSCut \citep{brasseur2019}, and performed photometry using the package {\tt lightkurve} \citep{lightkurve}. We used a 3x3 aperture centred on the star to avoid contamination by a bright ($V = 9.79$) star 2 arcmins away (which corresponds to only $\sim 6$~pixels in TESS). The background was estimated using the same aperture in a region with no stars. 
Using the VARTOOLS program \citep{HartmanBakos2016}, we performed a generalized Lomb-Scargle 
search \citep{ZechmeisterKuerster2009, Press1992} for periodic sinusoidal signals. In the periodogram (grey line in the 
top panel in Fig.~\ref{fig:tess}), we find the strongest signal at $P=0.069764\pm0.000005$\,d, with an associated false alarm probability
of $\log(FAP)=-152$. 
The error on the period was estimated by running a
Differential Evolution Markov Chain Monte Carlo (DEMCMC) routine
\citep{TerBraa+Cajok2006} employing the -nonlinfit command implemented in the
VARTOOLS program. 
The phase-folded, and phase-binned, TESS light curve is shown in red in the bottom panel of  Fig.~\ref{fig:tess}. The black line represents a fit of a harmonic series (equation 48 in \citealt{HartmanBakos2016}, also used for the DEMCMC), to the phase-folded light curve. The peak-to-peak amplitude (defined as 
the difference of the maximum and minimum of the fit) of the phase folded light curve is 23\,mmag. We note, however, that 
due to the long exposure time (about one-tenth of the period), neither the amplitude nor the shape of this phase-folded light 
curve can be considered as reliable. After whitening the light curve for this signal, no other significant peaks, no other significant peak remains in the periodogram (light blue line in the top panel in Fig.~\ref{fig:tess}).


The amplitude of the $P=0.069764$\,d peak is too high to be explained as a $g$-mode pulsation \citep{green2003}, although the period is in the correct range. Most likely the variability is explained by a spot on the surface of the star, driven by the magnetic field, leading to periodic variations in observed flux as the star rotates. The uneven minima might suggest that two magnetic dark spots are present rather than one, which would be consistent with a dipole magnetic field \citep[e.g.][]{Jagelka2019}. The non-sinusoidal shape of the phase-folded light curve (Fig.~\ref{fig:tess}) is also typical of rotational modulation \citep[e.g.][]{Angus2018}.
Assuming the radius $R = 0.1 R_{\odot}$ from the SED fit and rotational period of $P_{\rm rot}=0.069764$\,d, we derive a rotational velocity of $V_{\rm rot} = 73$\,km/s.



\section{Galactic orbit}
Based on the Gaia EDR3 data, we can calculate the Galactic orbit of \J. For this the {\sc galpy} \citep{Bovy2015} python package was used. The parameters used as input for the Galpy code are shown in Table\,\ref{tb:galorb}. They are all taken from Gaia EDR3, except the radial velocity. For the radial velocity, the weighted average of the RV measurements of the six spectra was taken. To calculate the errors, a Monte Carlo approach with 500 iterations was used. We find that \J\ has a Galactic orbit with a maximum height above the plane (Z$_{\rm max}$) of $688$ $\pm$ $288$ pc, a pericentre and apocentre distance (R$_{\rm per}$, R$_{\rm apo}$) of respectively $3.7$ $\pm$ $0.5$ kpc and $7.7$ $\pm$ $0.1$ kpc, an eccentricity (Ecc) of $0.36$ $\pm$ $0.05$ and an angular momentum of J$_{\rm z}$ = $1166$ $\pm$ $88$ kpc km/s. These parameters are also summarized in Table\,\ref{tb:galorb}.

When comparing to other hot subdwarf systems, from for example \citet{Luo2020}, \J\ would belong to the group of systems with relatively low J$_{\rm z}$ and above-average eccentricity (the average eccentricity in the sample of Luo is 0.23). It has similar kinematics as thick disc stars but is located close to the boundary between the thick and thin disk \citep{Pauli2006}. While \J\ lies on the edge of the J$_{\rm z}$-eccentricity regions occupied by the hot subdwarfs in the thin and thick discs, it is certainly not an outlier relative to either of these two populations.

One can link the kinematic properties of a system to its age. Thin disc stars are initially born on planar and circular orbits. Over time, interactions with different Galactic components (spiral arms, the bar, molecular clouds) make stellar orbits eccentric, induce radial migration and drive the orbits off-plane. Therefore, the present-day eccentricity and vertical extent of the orbit of \J\ may be linked to its age.

Asteroseismic observations combined with kinematics data show that stars typically found at Z-heights of about $500$ pc (which approximates the time-averaged absolute Z-location of the system) have ages between $2$ and $8$ Gyr \citep{Casagrande2016}. Taking into account that the progenitors of sdB stars need to evolve off the MS, this would correspond to progenitor's primary masses of between $0.9$ and $1.5$ M$_{\odot}$ \citep[see, e.g.][]{Vos2020}. Furthermore, \citet{Frankel2018}  showed that stars can migrate in radial direction by $4$ kpc on about $8$ Gyr timescale. Here the difference between R$_{\rm per}$ and R$_{\rm apo}$ of $4$ kpc can be taken as a proxy for this migration process. The corresponding migration timescale of $8$ Gyr is consistent with the age constraint based on the Z-location of the system. Summarised, the Galactic orbit is consistent with an interaction of two older stars (for example, a He double WD merger), as well as a different formation channel involving an initial primary with a mass of up to about $1.5$ M$_{\odot}$.

In the Gaia images, a nearby star at nearly the same distance as \J\ is visible. However, the two systems are likely not related. More information is given in Appendix\,\ref{ap:resolved-companion}.


\begin{table}[t]
    \centering
    \caption{Galactic orbit calculation of \J: input parameters for {\sc galpy} together with the resulting Galactic orbital parameters. The average values for general hot subdwarf stars are shown in the last column.}
    \label{tb:galorb}
    \begin{tabular}{lr@{ $\pm$ }lr@{ $\pm$ }l}
    \hline\hline
    \noalign{\smallskip}
    Parameter & \multicolumn{2}{c}{Value} & \multicolumn{2}{c}{sd average$^{\rm 1}$} \\
    \hline
    \noalign{\smallskip}
    \multicolumn{5}{c}{Input} \\
    \noalign{\smallskip}
    RA (dgr)         & \multicolumn{2}{l}{344.101358304}       \\
    Dec (dgr)        & \multicolumn{2}{l}{-59.170770517}       \\
    Distance (kpc)   & 0.635           & 0.013   \\
    PM RA (mas/yr)   & 13.772          & 0.017   \\ 
    PM Dec (mas/yr)  & -24.433         & 0.018   \\ 
    RV (km/s)        & 21              & 32      \\
    \noalign{\smallskip}
    \multicolumn{5}{c}{Galactic orbit} \\
    \noalign{\smallskip}
    Z$_{\rm max}$ (kpc)    & 0.688 & 0.288 &  1.14 & 0.70 \\ 
    R$_{\rm per}$ (kpc)    & 3.692 & 0.428 &  5.98 & 2.45 \\
    R$_{\rm apo}$ (kpc)    & 7.732 & 0.055 &  9.94 & 1.84 \\
    Ecc                    & 0.356 & 0.050 &  0.23 & 0.13 \\
    J$_{\rm z}$ (kpc km/s) & 1166  & 88    &  \multicolumn{2}{c}{/}\\
    \hline
    \end{tabular}
    \tablefoot{1: average values taken from \citet{Luo2020}.}
    
\end{table}

\section{Interpretation}

\begin{figure}
    \includegraphics[width=\linewidth]{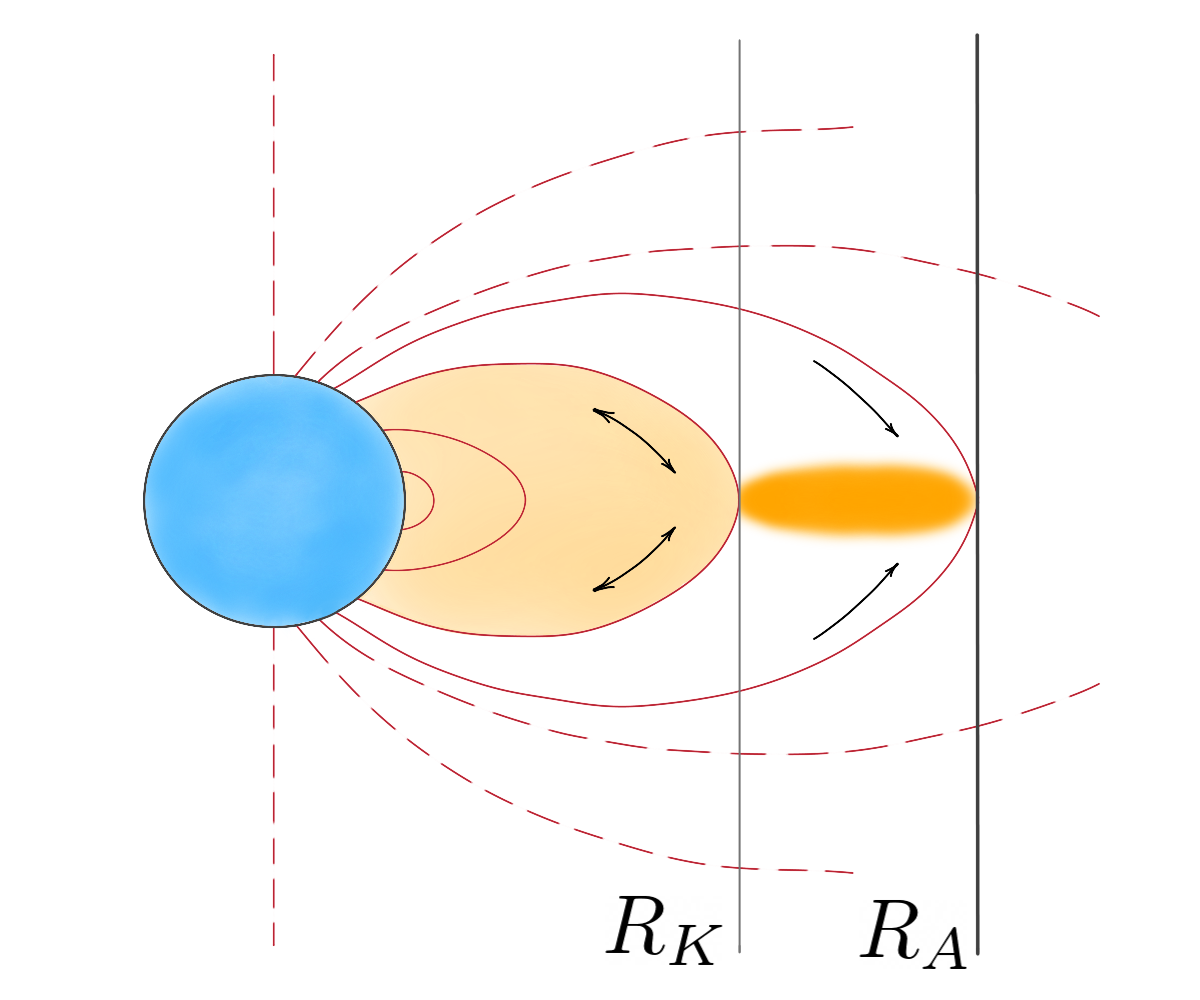}
    \caption{ Illustration for the structure of the system for the magnetic interpretation. The spinning subdwarf (in blue) generates a dipolar magnetic field around it, not necessarily aligned with the spin. The wind material in the innermost region, within Kepler radius $R_K$, has no centrifugal support and can fall freely back onto the star, thus forming a dynamical magnetosphere. The material between the Keplerian co-rotation radius $R_K$ and the Alfv\'{e}n radius $R_A$ has centrifugal support. It cannot fall back onto the star, thus forming a much denser rigidly rotating magnetosphere. The illustration is inspired by \citet{Petit+2013}.}\label{fig:MagScenario}
\end{figure}

\subsection{Magnetic fields} \label{s:magnetic_fields}

The triplet structure that is clearly visible in several hydrogen and helium lines immediately brings magnetic fields and Zeeman splitting to mind. This interpretation fits in with the expected evolution history of this system. As a single sdB formed by a merger, it is expected that \J ~would acquire a strong magnetic field,  generated through a dynamo process during the common-envelope evolution or the subsequent merger \citep[e.g.][]{Tout2008, GarciaBerro2012}.

\begin{figure*}
    \includegraphics[width=\linewidth]{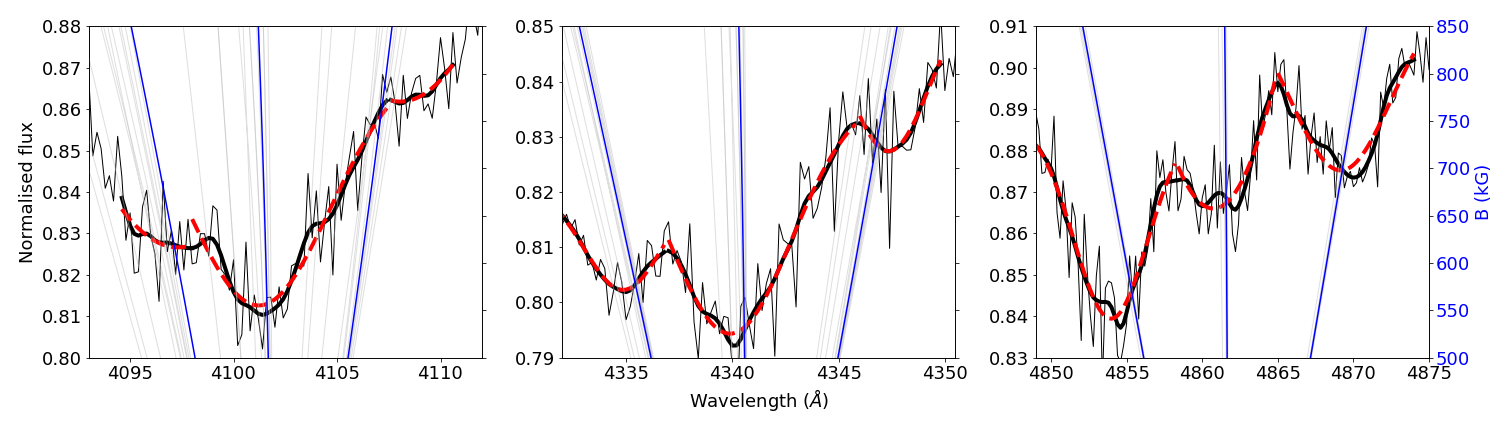}
    \caption{Magnetic field estimate for the last obtained X-SHOOTER spectrum. H$_{\delta}$, H$_{\gamma}$, and H$_{\beta}$ are shown from left to right. The observed spectrum is shown as a thin black line; a smoothed spectrum, restricted to the region where the Zeeman components were fitted, is shown as a thicker line. The fits to each Zeeman component are shown as red dashed lines. The models of \citet{Schimeczek2014} are shown in grey, with the averaged separation including higher-order terms shown in blue, and the corresponding field strength shown on the right-hand side. The estimate for this spectrum was $660\pm62$~kG.}\label{fig:mag_field}
\end{figure*}

We have applied a method similar to that of \citet{Kepler2013} to estimate the field strength necessary to produce the observed line splitting. The method relies on the fact that, for magnetic fields in the range $10\,{\rm kG} - 2\,{\rm MG}$, the observed line shift $\Delta \lambda $ caused by a field $B$ to the hydrogen lines can be approximated in the first order by
\begin{equation}
    \delta\lambda = \pm\,4.67 \cdot 10^{-7}\lambda^{2}B,
\end{equation}
where the wavelength is measured in $\AA$ and the magnetic field in MG. To account for contributions of higher-order terms, we have utilised the models of \citet{Schimeczek2014} (see their figure 5) to estimate the averaged component separation predicted by the models. To calculate the observed separation for our obtained spectra, Gaussian lines were fitted to each Zeeman component. We only used the lines H$_{\beta}$, H$_{\gamma}$ and H$_{\delta}$, as for higher-order lines the triplet structure is not apparent even for low fields \citep[see, e.g. figure 5 of][]{Kepler2013}, and H$_{\alpha}$ is seen in emission. Moreover, we only applied this method to the spectra in which the three components could be identified for these lines. Depending on field structure and orientation, one or more components can be suppressed. Our method is illustrated in Fig.~\ref{fig:mag_field}. For each spectrum, we searched for the field strength whose predicted separation could better explain the observed spectrum by minimising the difference between observed and predicted separation for the three lines simultaneously. To estimate uncertainties, we drew flux values a thousand times, assuming a 10\% uncertainty on the observed values, and repeated the estimate for each of the simulated spectra. The results are shown in Table~\ref{tb:field_strength}. Assuming that the field does not change significantly over time, which seems to be suggested by our consistent estimates, the average field is $ 656\pm51$~kG.\\

But how can we now understand the H\,$\alpha$ line profile in view of this high magnetic field strength? Emission in H\,$\alpha$ is not an atypical phenomenon amongst magnetic stars. Magnetically active (sub-)giants, for example, show chromospheric emission lines in H\,$\alpha$, but at the same time also in the cores of the \ion{Ca}{ii} H and K lines, as well as sometimes other lines in the optical or ultraviolet \citep{Wilson1963, Wilson1968, GrayCorbally2009}. For some of these stars, time-variability in the chromospheric H\,$\alpha$ emission, that is not correlated to the rotation period, has also been reported, though, the exact origin of the variability is not yet understood (e.g., \citealt{Dorren+1984, Vida+2015, Kovari+2019, Werner+2020}).

There is also a small group of three cool (effective temperatures between 7500\,K and 7865\,K), magnetic and apparently single white dwarfs known that exhibit Zeeman-split Balmer emission lines \citep{GreensteinMcCarthy1985, Reding+2020, Gaensicke+2020}.
It is thought that a conductive planet in a close orbit around these stars could result in the generation of electric currents that heat the regions near the magnetic poles of the white dwarf. The planet, in this case, would have formed in a metal-rich debris disc that was left over by a double white dwarf merger that could have produced the magnetic white dwarf \citep{Li+1998, Wickramasinghe+2010}. However, in contrast to our star,
the emission lines in these white dwarfs are not only seen in H\,$\alpha$ but also H\,$\beta$ and are triple-peaked instead of single/double-peaked.

Last but not least, for magnetic O- and B- type main sequences stars that host a wind-fed, co-rotating, circumstellar magnetosphere, emission in  H$\alpha$ is the primary visible magnetospheric diagnostic. In Fig.\,\ref{fig:MagScenario}, we show such model applied to our system. In slowly rotating stars, the material persists within the magnetosphere only over the free-fall timescale, and is pulled back onto the star by
gravity (so-call dynamical magnetosphere, \citealt{Landstreet+Borra1978, Ud-Doula+2002, Petit+2013}). However, if the star is rapidly rotating or the magnetic field strength is high enough, the co-rotating material in the magnetosphere can reach high enough rotational velocities so that the gravitational infall can be prevented. This is the case when the Alfv\'{e}n radius $R_A$, which characterises the maximum height of closed magnetic loops, exceeds the Keplerian co-rotation radius, $R_K$ (point of balance between
gravitational and centrifugal force). 
While below $R_K$, the star retains a dynamical magnetosphere, above $R_K$ and extending to $R_A$, a so-called centrifugal magnetosphere forms. Herein, the trapped wind material accumulates into a relatively dense, stable and long-lived ``rigidly rotating magnetosphere'' (RRM, \citealt{Townsend+Owocki2005, Townsend+2007}). According to the RRM model, the distribution of the material then depends on the tilt, $\beta$, of the magnetic axis with respect to the rotational axis of the star. While for $\beta=0$\textdegree\ a continuous torus in the magnetic equatorial plane forms, two distinct plasma clouds are expected near the intersections of the magnetic and rotational equatorial planes for $\beta=90$\textdegree. The typical RRM geometry, thus, produces a
double-humped emission profile, when the circumstellar magnetosphere is seen face on.
Since close to $R_K$, the magnetosphere has its highest density, also the H$\alpha$ emission is found to peak close to $R_K$ (typically around $1.25\times R_K$, \citealt{Shultz+2020}). The remaining shape of the H$\alpha$ emission then depends on the geometry of the RRM. Non-eclipsing stars with small $\beta$ show emission at all velocities across the line profile, whereas non-eclipsing stars with large $\beta$ will display emission only outside of $\pm R_K$ at maximum emission. The emission line profile is modulated by the rotation of the object. As the projected distance of the clouds from the star decreases, and -- at the same time -- the projected area of the clouds becomes smaller when changing from face-on to edge-on, the H\,$\alpha$ emission bumps decrease in strength \citep{Shultz+2020}.

Since also in our star, we detect this time-variable, double-humped H$\alpha$ emission profile, the RRM model appears attractive. The multi-component absorption features seen for the CaII and OI originating in the circumstellar material (Fig.~\ref{fig:spectral_trails_non_phot}) could also be explained by the distribution of the material in the magnetosphere. Magneto-hydrodynamic simulation studies (e.g.,
\citealt{Ud-Doula+2002, Ud-Doula+2008}) show that in case of a large-scale, dipole magnetic field, a magnetosphere forms when the wind magnetic confinement parameter ($\eta_{\mathrm{\star}}$) is larger than one:
\begin{equation*}
\eta_{\mathrm{\star}}=\frac{B_{\mathrm{eq}}^2 R_{\mathrm{\star}}^2}{\dot{M}_{B=0} v_\infty} > 1.
\end{equation*}
Here $B_{\mathrm{eq}}=B_d/2$ is the field strength at the magnetic equatorial surface radius, $R_{\mathrm{\star}}$, and \Mdot$_{B=0}$ and $v_\infty$ are the fiducial mass-loss rate and terminal wind speed that the star would have in the absence of any magnetic field (all in cgs units). Assuming a typical mass loss rate for an sdB star of $10^{-11.5}$\Msol/yr \citep{VinkCassisi2002}, $v_\infty=v_{esc}=1338$km/s (assuming $M=0.47$\,\Msol, and $R=0.1$\,\Rsol, \citealt{Hamann+1981, Howarth1987}), and $R=0.1$\,\Rsol, we find that in case of \J\ a magnetosphere can already form for $B_{eq}\gtrapprox24$\,G. This is many orders of magnitude below of what we find above, thus, the requirement for a centrifugal magnetosphere ($R_A>R_K$) can be easily fulfilled. Assuming $M=0.47$\,\Msol, and $R=0.1$\,\Rsol, we find $R_K=5.5\,R_{\mathrm{\star}}$ assuming the $0.07$\,d period observed in the TESS light curve is the rotational period of the star. The Alfv\'{e}n radius, $R_A$, can be estimated from the wind magnetic confinement parameter, $\eta_*$, via $R_A/R_{\mathrm{\star}}\approx0.3\,(\eta_*+0.4)^{1/4}$ \citep{Ud-Doula+2008}.
Here, we find $R_A=118.6\,R_{\mathrm{\star}}$, but it should be noted that the exact 
value depends on the mass loss rate and terminal wind velocity, which we can only estimate. Moreover, additional circumstellar material might be present as a result of the possible merger, which we do not take into account here. What can, however, be taken away from this is that a centrifugal magnetosphere can be expected\footnote{Observationally, centrifugal magnetospheres are detected in stars with $\log(R_A/R_K) >0.7$ \citep{Shultz+2020}}.

Since the rigid-body rotation of the circumstellar magnetosphere implies, that the line of sight velocity, v, is directly proportional to the projected distance from the star ($v/(v_{rot}\,sin\,i)=r/R_{\mathrm{\star}}$), one can in principle test the circumstellar magnetosphere scenario with the H$\alpha$ line profile directly \citep{Shultz+2020}, as the emission peaks should occur around $R_K$ (see above). We find that the observed emission peaks of the H\,$\alpha$ line in \J\ would be located at $R_K$ if we assume for the 0.07\,d period an inclination angle of $i\approx20$\textdegree\ (blue dashed-dotted lines in Fig.\ref{fig:spectral_trails_phot}). Unfortunately, due to the lack of any photospheric metal lines and the high magnetic field, it is not possible 
to measure $v_{rot}\,sin\,i$, plus the lack of the knowledge of the stellar mass adds 
another uncertainty when calculating $R_K$. In addition, we note that if the 0.07\,d period is indeed the rotational period, then the spectra should suffer considerably from rotational smearing due to the long exposure time (one quarter of the period). Hence the line profile shapes may not be considered as reliable. Thus, based on the current data, it is not possible to entirely confirm the RRM model.

\begin{table}[t]
    \centering
    \caption{Field estimates for the three spectra in which three Zeeman components can be identified for the H$_{\beta}$, H$_{\gamma}$ and H$_{\delta}$ lines. The magnetic field strength is estimated using the method of \citep{Kepler2013}. See Sect.\,\ref{s:magnetic_fields} for details.}
    \begin{tabular}{lll}
    \hline\hline
    \noalign{\smallskip}
    MJD & Instrument & $B$ (MG) \\
    \hline
    \noalign{\smallskip}
    58731.004125    &  UVES      & $640\pm14$ \\
    58773.062886    &  X-SHOOTER & $670\pm59$ \\ 
    58804.105317    &  X-SHOOTER & $660\pm62$ \\ 
    \hline
    \end{tabular}
    \label{tb:field_strength}
\end{table}

\subsection{Circumstellar material (non-magnetic scenario)} \label{s:disk}

Complicated spectral line profiles may also have another explanation that does not require magnetic fields. In this section, we investigate the idea that the spectral lines are shaped by circumstellar material (CM). Such material is often present in interacting binaries or hot stars and might be present in our system too. It gives rise to emission lines of quite complicated shape which may be superposed on the absorption lines originating from the stellar atmosphere. The result would be even more complicated spectral line profiles. In this scenario double absorption
 like in the H$_{\alpha}$ line would not be composed of two absorption lines but from a single broad absorption line (from the stellar atmosphere) filled up by more narrow central emission from the CM. 
 Lines with triple absorption could be understood as a single broad absorption
 from the photosphere filled in by a double peak emission from CM. 

The most natural form of circumstellar material in a binary system or in a merger of two stars is probably an accretion disk. Typically such a disk gives rises to a double-peaked emission (unless seen pole-on).
Thus an inclined disk might explain triple absorption profile seen best in H$\beta$.
To demonstrate the idea that such line profiles may be due to CM, we performed a synthetic spectra calculation. As this analysis aims to be a qualitative one and not a quantitative fit to the spectra, we use a master spectrum for the comparison of the models with the observations. This master spectrum is obtained by summing all three X-SHOOTER spectra without any RV corrections for stellar motion. The spectra were then normalised.

\begin{figure}
    \includegraphics[width=6.2cm,angle=-90]{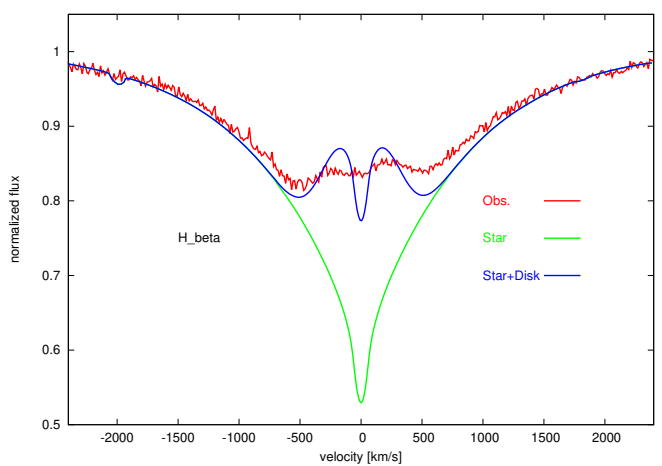}
    \caption{The synthetic spectra of the H$_{\beta}$ line compared with the observations. The theoretical spectrum of the star without the disk is shown in green. The theoretical spectrum of the star + disk is shown in blue.
    The observed spectrum is shown in red. More details may be found in Section\,\ref{s:disk}.} \label{fig:disk_hbeta}
\end{figure}

As a first step we calculated the stellar atmosphere model using the TLUSTY code \citep{Hubeny1995}. These are 1D Non-Local-Thermodynamic-Equilibrium (NLTE) atmosphere models. The spectra emerging from these atmosphere models were calculated using the code {\sc synspec} \citep{Hubeny2017}. We assumed T$_{\rm eff} = 26000$ K, $\log{g} = 6.1$ [cgs], and solar chemical composition. 
Such synthetic spectra of H$_{\beta}$ line are shown in Fig. \ref{fig:disk_hbeta}. One can see very strong, deep and relatively sharp absorption
originating from the stellar atmosphere. The major challenge of this model is filling this profile with emission. This intrinsic spectrum of the star was used as a boundary condition to calculate the spectra of the star and CM, for which we used the {\sc shellspec} code \citep{Budaj2004}. 
It is designed to calculate light curves and spectra of interacting binaries embedded in a 3D moving CM, assuming local thermodynamic equilibrium (LTE) and optically thin scattering. The assumed stellar mass, radius, and projected equatorial rotation velocity are M = 0.47 M$_{\odot}$, R = 0.1 R$_{\odot}$, and $v\sin{i} = 70$ km/s, respectively. Quadratic limb darkening coefficients for the star from \citet{Claret2000} were assumed. The chemical composition of the CM was identical to that of the star. 
CM had a form of an accretion disk. Synthetic spectra of the most interesting and most complicated H$\beta$ line are also shown in Fig. \ref{fig:disk_hbeta}.
One can clearly see a double peak emission from the disk filling in the central part of the absorption from the photosphere. The result might look like a triple absorption.

The properties of the CM are described below.
The disk was modelled using an object called NEBULA in the {\sc shellspec} code.
It is a flared disk characterised by its inner, $R_{in}$, and outer, $R_{out}$, radius, and an inclination $i$. Its density is decreasing in the radial direction as a power law, $\rho(r)=\rho_{\rm in}(r/r_{\rm in})^{\rho_{exp}}$,
and in the vertical direction as a Gaussian. It is characterised by the density at the inner radius $\rho_{in}$ and exponent $\rho_{exp}$.
We assumed $\rho_{exp}=-1$ based on \cite{Budaj2005}. 
The temperature, similarly, has a radial power-law dependence characterised by $T_{in}$ and exponent $T_{exp}$. 
The velocity field is Keplerian. In reality, it may be much more complicated.
The disk may have a radial inflow component
and is also often accompanied by winds, jets, or other outflows.
That is why we also introduce a simple parameter called 'turbulence', $v_{T}$.
Electron number densities are calculated from the density, temperature, and chemical composition assuming LTE.
Values of all these parameters are summarised in Table \ref{tb:cm}.

\begin{table}[t]
    \caption{Properties of the circumstellar medium (CM) used in the {\sc shellspec} model. The CM is assumed to have disk geometry, limited in extent by $R_{in}$ and $R_{out}$. The radial density and temperature profiles are given as power laws, $\rho(r)=\rho_{\rm in}(r/r_{\rm in})^{\rho_{exp}}$ and $T(r)=T_{\rm in}(r/r_{\rm in})^{T_{exp}}$. $v_{T}$ is the turbulence line broadening parameter.} \label{tb:cm}
    
    \centering
    \begin{tabular}{lc}
    \hline\hline
    \noalign{\smallskip}
    \multicolumn{2}{c}{Disk} \\
    \hline
    \noalign{\smallskip}
    $R_{in} [R_{\odot}]$    &  0.7                \\
    $R_{out} [R_{\odot}]$   &  3.3                \\
    $\rho_{in}$ [cgs]       &  6.2\,$10^{-13}$    \\
    $\rho_{exp}$ []         &  -1.                \\
    $T_{in}$  [K]           &  17000              \\
    $T_{out}$  [K]          &  7800               \\
    $T_{exp}$ []            & -0.5                \\
    $v_{T}$ [km\,s$^{-1}$]  & 230                 \\
    i [deg]                 & 70                  \\
    \hline
    \end{tabular}
    
\end{table}

We can conclude that CM material might explain the complicated shapes of spectral lines we observe in this star. However, these calculations should be understood only as a demonstration of the effect and disk parameters (mainly densities) represent
rather a lower limit. In reality, the geometry, velocity field and behaviour of state quantities of the CM may be much more complicated than our simple disk model. Their effect will be mainly to smooth the ideal theoretical line profile. We experimented with dozens of other models of CM like shells, disks, temperature inversions, and many of them produce a similar outcome, i.e. triple absorption profiles. The most difficult problem is to fill in the sharp central absorption peak.

The advantage of this model is that it has the potential also to
explain the IR excess observed in the SED, as well as the emission seen in other H and He lines in the spectra. The change in the line profiles that is shown in Fig.\,\ref{fig:spectral_trails_phot} might be explained by the disk
model as well. Accretion structures are not necessarily stable and can change over time, causing changes in the line profiles. If a magnetic field is present, this can also affect the structure and cause variability. Furthermore, if there would be a secondary body present in the system, it can cause precession of the disk which would cause changes in the emission cores of the H and He lines. This body would have to be much closer than the nearby companion described in Appendix\,\ref{ap:resolved-companion}.

\subsection{CE outcome versus merger}
The three formation channels for hot subdwarf stars are stable RLOF, CE ejection and a merger. Based on the observations we can exclude two of these channels with a very high likelihood.

All known wide sdB binaries formed through the stable RLOF channel have FGK type companions \citep[e.g.][]{Vos2017}. These companions can be clearly seen in high-resolution spectroscopy. There is no sign of such a companion in \J, and thus this scenario can be excluded. There are then two possibilities left. \J\ can be a close binary with a dM or WD companion, or it can be a single merger product.

\begin{figure}
    \includegraphics[width=\linewidth]{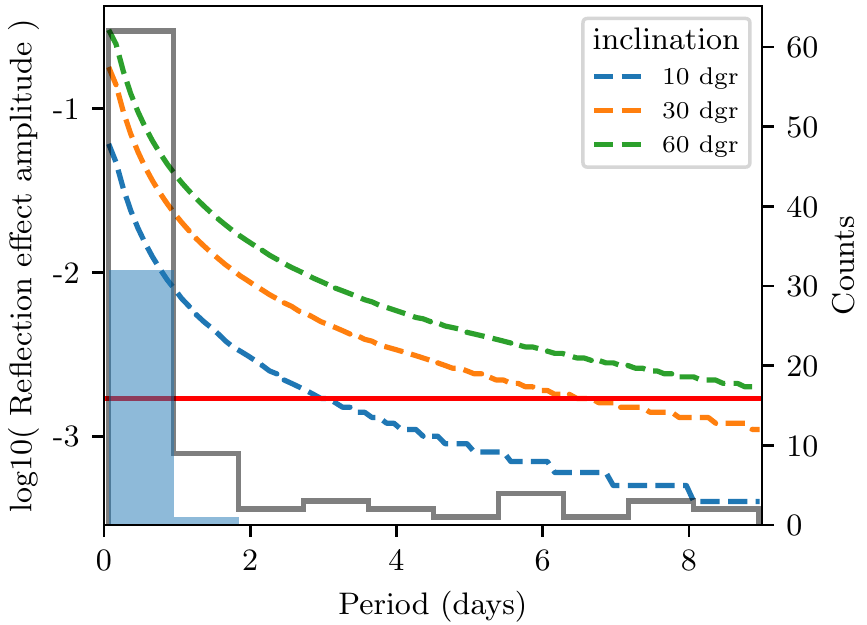}
    \caption{The expected reflection effect for a typical dM companion at different inclination angles as a function of orbital period. In full red line, the detection limit of TESS is given. On the right axis, the known period distribution of sdB+dM binaries is shown in blue. In black, the period distribution of sdB binaries with an unknown companion (low mass MS or WD) is shown. The period distributions were taken from \citet{Kupfer2015}.}
    \label{fig:reflection_effect_amp_sdB+dM}
\end{figure}

In the case that \J\ would be an sdB+dM binary, we can compare it to the known population of sdB+dM binaries. These systems have all short orbits, with their period distribution peaking at less than a day. Using the {\sc lcurve} package \citep[App A.]{Copperwheat2010}, we have computed the expected amplitude for the reflection effect in sdB+dM binaries for a typical dM companion at different orbital periods and inclinations angles (see Fig.\,\ref{fig:reflection_effect_amp_sdB+dM}). From this figure, it is clear that such a binary would be detected in the TESS light curve for orbital periods up to $\sim 8$ days. Since these systems are nearly never found at orbital periods larger than a few days, we can conclude that the sdB+dM possibility is very unlikely. 

The above considerations leave the sdB+WD possibility. Such systems do not show a significant reflection effect in the light curves but can show ellipsoidal modulations. {\sc lcurve} models show that such effects would be detectable for typical WD companions for orbital periods up to $\sim 0.3$ days. As sdB+WD binaries are observed on longer orbital periods than that, the light curve alone is not sufficient to exclude this possibility. 

To further judge the sdB+WD option, the RVs derived from the spectra can be used. Known sdB+WD systems have short orbital periods and thus high RV variations. The RVs that are derived in Sect.\,\ref{s:rv} show possible variations of up to $\sim30$ km/s. This would correspond to sdB+WD systems on orbital period $> 5$ days, which is very exceptional for sdB+WD systems \citep{Kupfer2015, Prince2019}. The RV analysis used the wings of the H lines, which originate from the central star, and are likely not significantly influenced by the disk.

Based on the limits obtained from the light curve and the spectra, we can with high certainty say that \J\ is a single sdB and thus a merger product. The derived properties suggest that the object is in a core-He burning phase. For typical double-degenerate mergers, this phase is only reached long after the merger ($\sim 10^6$~yr), at which point all the circumstellar material resulting from the merger should have been lost. Additionally, He-core burning products of double degenerate mergers are expected to show higher temperatures and be H-deficient, which is not the case here \citep{Dan2014,Schwab2018}. Therefore, a WD+MS merger is a more likely scenario for the formation of \J\, as in this case the He-core burning phase can occur at lower effective temperature, and high amounts of hydrogen are still expected \citep{Zhang2017}.

In case that \J\ would indeed be a magnetic system there is an extra argument to be made against the CE formalism. While the dynamo actions arising during the CE phase can indeed cause magnetic fields, models show that these fields are weak and don't last long after the CE ejection \citep[see e.g.][]{Potter2010}.

\section{Discussion and conclusions}
\J\ is a hot subdwarf star with very shallow, variable, multi-peaked H and He lines, with an H$_{\alpha}$ emission, IR excess, photometric variability, and possibly with a high rotational velocity. Based on the multi-band photometry, time resolved spectroscopy, Gaia astrometry and the TESS light curve, we have found two possible interpretations of the spectrum. The first one is that the multi-peaked spectral features are caused by the existence of a magnetic field of about 650 kG. The IR excess in the SED and the emission core in the H$_{\alpha}$ line can then be caused by the formation of a magnetosphere. Similarly, this would explain the time variation of the spectral lines. A second possibility is the existence of a circumstellar disk that would explain both the IR excess and the special line shapes. Changes in the line shapes can be explained by variations in the circumstellar material or a precessing disk. 

The major problem with this star is that the observed hydrogen lines are too shallow. Helium lines are equally shallow and broad compared with theoretically expected lines originating from the stellar photosphere. 
Some other hot subdwarf stars show a relatively narrow emission in $H{\alpha}$. This, however, happens at much hotter temperatures and is caused by NLTE effects that affect the level populations of the hydrogen atom \citep[e.g.][]{Rauch2010, Reindl2014, Latour2015, Dorsch2020}. Thus we can exclude that this is the reason of the observed emission in $H{\alpha}$ and shallowness of other H, He lines.

All hot subdwarf stars are formed through binary interaction, whether it is stable RLOF, a CE ejection phase or a merger. In this case, the observations point strongly to a post-merger single sdB. A CE ejection resulting in a close binary with a WD companion is possible but very unlikely. The presence of circumstellar material in the system and the line variations would indicate that we are observing this system very closely after the interaction phase. Both of our interpretations of the observations are consistent with the merger scenario, as a magnetic field could be instilled in the final product, and a significant amount of mass would end up around the sdB.

Regardless of whether the observed features are caused by magnetism or circumstellar matter, \J\ is a very interesting system that could provide insight in the early dawn after a binary merger phase. It has the potential to solve several outstanding problems in the physical explanation of this phase, including the discrepancy between the predicted high rotational velocities for post merger products versus the observed low rotation rates of single sdBs, or the predictions that mergers can instill magnetic fields in their products. If \J\ turns out to be a a magnetic sdB, then it could be a long sought immediate ancestor of strongly magnetic WDs. This could provide vital clues to understand the magnetic field evolution across the Hertzsprung Russell diagram.

The likely origin of \J\ is the CE evolution of an RGB and a He-WD. The CE episode would either have led to a merger within the red giant envelope, between the He-RG core and the He-WD, or to a short-period He-WD binary which later merges due to gravitational wave emission \citep{Han2002}. In this case, the observed properties of \J\ may be explained by it being a particularly young member of the class of single sdBs.

The two interpretations offered here are not mutually exclusive and are neither the only possible explanations of this system, although in our opinion they are the most likely. It is perfectly possible that both a circumstellar disk and a magnetic field are present in the system. The spectrum of this system contains multiple components which with the currently available observations are impossible to disentangle. Further investigation of, for example, spectropolarimetry can confirm if a magnetic field is really present in the system. Time resolved spectroscopy will allow the investigation of the line profile variations. Observations in the UV or even in X-ray domain with eRosita would be valuable to constrain models of the system as for example X-ray flares may be expected \citep[e.g.][]{Groote2004}. On the other end of the spectrum, observations in the far IR with for example ALMA would provide clues to structure of the gas and dust present in the system.

\begin{acknowledgements}
The authors would like to thank Uli Heber for his very helpful referee report.
This work was supported by a fellowship for postdoctoral researchers from the Alexander von Humboldt Foundation awarded to JV.
IP was partially funded by the Deutsche Forschungsgemeinschaft under grant GE2506/12-1.
JB was supported by the VEGA 2/0031/18 and by the Slovak Research and Development Agency under the contract No. APVV-20-0148.
V.S. is supported by the Deutsche Forschungsgemeinschaft (DFG) through grant GE 2506/9-1
M.U. acknowledges financial support from CONICYT Doctorado Nacional in the form of grant number No: 21190886 and the ESO studentship program.
Based on observations collected at the European Southern Observatory under ESO programmes 0101.D-0440, 0103.D-0129 and 0104.D-0596. Based in part on observations obtained at the SOAR telescope, which is a joint project of the Minist\'{e}rio da Ci\^{e}ncia, Tecnologia e Inova\c{c}\~{o}es (MCTI/LNA) do Brasil, the US National Science Foundation’s NOIRLab, the University of North Carolina at Chapel Hill (UNC), and Michigan State University (MSU). 
This work has made use of data from the European Space Agency (ESA) mission {\it Gaia} (\url{https://www.cosmos.esa.int/gaia}), processed by the {\it Gaia} Data Processing and Analysis Consortium (DPAC, \url{https://www.cosmos.esa.int/web/gaia/dpac/consortium}). Funding for the DPAC has been provided by national institutions, in particular the institutions participating in the {\it Gaia} Multilateral Agreement.
This research made use of Astropy, a community-developed core Python package for Astronomy \citep{Astropy2013}.
\end{acknowledgements}

\bibliographystyle{aa}
\bibliography{bibliography}

\begin{thebibliography}{105}
\expandafter\ifx\csname natexlab\endcsname\relax\def\natexlab#1{#1}\fi

\bibitem[{{Angus} {et~al.}(2018){Angus}, {Morton}, {Aigrain}, {Foreman-Mackey},
  \& {Rajpaul}}]{Angus2018}
{Angus}, R., {Morton}, T., {Aigrain}, S., {Foreman-Mackey}, D., \& {Rajpaul},
  V. 2018, \mnras, 474, 2094

\bibitem[{{Astropy Collaboration} {et~al.}(2013){Astropy Collaboration},
  {Robitaille}, {Tollerud}, {Greenfield}, {Droettboom}, {Bray}, {Aldcroft},
  {Davis}, {Ginsburg}, {Price-Whelan}, {Kerzendorf}, {Conley}, {Crighton},
  {Barbary}, {Muna}, {Ferguson}, {Grollier}, {Parikh}, {Nair}, {Unther},
  {Deil}, {Woillez}, {Conseil}, {Kramer}, {Turner}, {Singer}, {Fox}, {Weaver},
  {Zabalza}, {Edwards}, {Azalee Bostroem}, {Burke}, {Casey}, {Crawford},
  {Dencheva}, {Ely}, {Jenness}, {Labrie}, {Lim}, {Pierfederici}, {Pontzen},
  {Ptak}, {Refsdal}, {Servillat}, \& {Streicher}}]{Astropy2013}
{Astropy Collaboration}, {Robitaille}, T.~P., {Tollerud}, E.~J., {et~al.} 2013,
  \aap, 558, A33

\bibitem[{{Belokurov} {et~al.}(2020){Belokurov}, {Penoyre}, {Oh}, {Iorio},
  {Hodgkin}, {Evans}, {Everall}, {Koposov}, {Tout}, {Izzard}, {Clarke}, \&
  {Brown}}]{Belokurov2020}
{Belokurov}, V., {Penoyre}, Z., {Oh}, S., {et~al.} 2020, \mnras, 496, 1922

\bibitem[{{Bovy}(2015)}]{Bovy2015}
{Bovy}, J. 2015, \apjs, 216, 29

\bibitem[{{Brasseur} {et~al.}(2019){Brasseur}, {Phillip}, {Fleming},
  {Mullally}, \& {White}}]{brasseur2019}
{Brasseur}, C.~E., {Phillip}, C., {Fleming}, S.~W., {Mullally}, S.~E., \&
  {White}, R.~L. 2019, {Astrocut: Tools for creating cutouts of TESS images}

\bibitem[{{Budaj} \& {Richards}(2004)}]{Budaj2004}
{Budaj}, J. \& {Richards}, M.~T. 2004, Contributions of the Astronomical
  Observatory Skalnate Pleso, 34, 167

\bibitem[{{Budaj} {et~al.}(2005){Budaj}, {Richards}, \& {Miller}}]{Budaj2005}
{Budaj}, J., {Richards}, M.~T., \& {Miller}, B. 2005, \apj, 623, 411

\bibitem[{{Casagrande} {et~al.}(2016){Casagrande}, {Silva Aguirre},
  {Schlesinger}, {Stello}, {Huber}, {Serenelli}, {Sch{\"o}nrich}, {Cassisi},
  {Pietrinferni}, {Hodgkin}, {Milone}, {Feltzing}, \&
  {Asplund}}]{Casagrande2016}
{Casagrande}, L., {Silva Aguirre}, V., {Schlesinger}, K.~J., {et~al.} 2016,
  \mnras, 455, 987

\bibitem[{{Claret}(2000)}]{Claret2000}
{Claret}, A. 2000, \aap, 363, 1081

\bibitem[{{Clemens} {et~al.}(2004){Clemens}, {Crain}, \&
  {Anderson}}]{2004SPIE.5492..331C}
{Clemens}, J.~C., {Crain}, J.~A., \& {Anderson}, R. 2004, in Society of
  Photo-Optical Instrumentation Engineers (SPIE) Conference Series, Vol. 5492,
  Ground-based Instrumentation for Astronomy, ed. A.~F.~M. {Moorwood} \&
  M.~{Iye}, 331--340

\bibitem[{{Copperwheat} {et~al.}(2010){Copperwheat}, {Marsh}, {Dhillon},
  {Littlefair}, {Hickman}, {G{\"a}nsicke}, \& {Southworth}}]{Copperwheat2010}
{Copperwheat}, C.~M., {Marsh}, T.~R., {Dhillon}, V.~S., {et~al.} 2010, \mnras,
  402, 1824

\bibitem[{{Dan} {et~al.}(2014){Dan}, {Rosswog}, {Br{\"u}ggen}, \&
  {Podsiadlowski}}]{Dan2014}
{Dan}, M., {Rosswog}, S., {Br{\"u}ggen}, M., \& {Podsiadlowski}, P. 2014,
  \mnras, 438, 14

\bibitem[{{Dorren} {et~al.}(1984){Dorren}, {Guinan}, \& {McCook}}]{Dorren+1984}
{Dorren}, J.~D., {Guinan}, E.~F., \& {McCook}, G.~P. 1984, \pasp, 96, 250

\bibitem[{{Dorsch} {et~al.}(2020){Dorsch}, {Latour}, {Heber}, {Irrgang},
  {Charpinet}, \& {Jeffery}}]{Dorsch2020}
{Dorsch}, M., {Latour}, M., {Heber}, U., {et~al.} 2020, \aap, 643, A22

\bibitem[{{Elkin}(1996)}]{Elkin1996}
{Elkin}, V.~G. 1996, \aap, 312, L5

\bibitem[{{Ferrario} {et~al.}(2015){Ferrario}, {de Martino}, \&
  {G{\"a}nsicke}}]{Ferrario2015}
{Ferrario}, L., {de Martino}, D., \& {G{\"a}nsicke}, B.~T. 2015, \ssr, 191, 111

\bibitem[{{Frankel} {et~al.}(2018){Frankel}, {Rix}, {Ting}, {Ness}, \&
  {Hogg}}]{Frankel2018}
{Frankel}, N., {Rix}, H.-W., {Ting}, Y.-S., {Ness}, M., \& {Hogg}, D.~W. 2018,
  \apj, 865, 96

\bibitem[{Gaia~Collaboration(2020)}]{GaiaEDR3}
Gaia~Collaboration, e.~a. 2020, \aap, in prep.

\bibitem[{{G{\"a}nsicke} {et~al.}(2020){G{\"a}nsicke}, {Rodr{\'\i}guez-Gil},
  {Gentile Fusillo}, {Inight}, {Schreiber}, {Pala}, \&
  {Tremblay}}]{Gaensicke+2020}
{G{\"a}nsicke}, B.~T., {Rodr{\'\i}guez-Gil}, P., {Gentile Fusillo}, N.~P.,
  {et~al.} 2020, \mnras, 499, 2564

\bibitem[{{Garc{\'\i}a-Berro} {et~al.}(2012){Garc{\'\i}a-Berro},
  {Lor{\'e}n-Aguilar}, {Aznar-Sigu{\'a}n}, {Torres}, {Camacho}, {Althaus},
  {C{\'o}rsico}, {K{\"u}lebi}, \& {Isern}}]{GarciaBerro2012}
{Garc{\'\i}a-Berro}, E., {Lor{\'e}n-Aguilar}, P., {Aznar-Sigu{\'a}n}, G.,
  {et~al.} 2012, \apj, 749, 25

\bibitem[{{Gray} \& {Corbally}(2009)}]{GrayCorbally2009}
{Gray}, R.~O. \& {Corbally}, Christopher, J. 2009, {Stellar Spectral
  Classification}

\bibitem[{{Green} {et~al.}(2003){Green}, {Fontaine}, {Reed}, {Callerame},
  {Seitenzahl}, {White}, {Hyde}, {{\O}stensen}, {Cordes}, {Brassard}, {Falter},
  {Jeffery}, {Dreizler}, {Schuh}, {Giovanni}, {Edelmann}, {Rigby}, \&
  {Bronowska}}]{green2003}
{Green}, E.~M., {Fontaine}, G., {Reed}, M.~D., {et~al.} 2003, \apjl, 583, L31

\bibitem[{{Greenstein} \& {McCarthy}(1985)}]{GreensteinMcCarthy1985}
{Greenstein}, J.~L. \& {McCarthy}, J.~K. 1985, \apj, 289, 732

\bibitem[{{Groote} \& {Schmitt}(2004)}]{Groote2004}
{Groote}, D. \& {Schmitt}, J.~H.~M.~M. 2004, \aap, 418, 235

\bibitem[{{Hall} \& {Jeffery}(2016)}]{Hall2016}
{Hall}, P.~D. \& {Jeffery}, C.~S. 2016, \mnras, 463, 2756

\bibitem[{{Hamann} {et~al.}(1981){Hamann}, {Gruschinske}, {Kudritzki}, \&
  {Simon}}]{Hamann+1981}
{Hamann}, W.~R., {Gruschinske}, J., {Kudritzki}, R.~P., \& {Simon}, K.~P. 1981,
  \aap, 104, 249

\bibitem[{{Han} {et~al.}(2003){Han}, {Podsiadlowski}, {Maxted}, \&
  {Marsh}}]{Han2003}
{Han}, Z., {Podsiadlowski}, P., {Maxted}, P.~F.~L., \& {Marsh}, T.~R. 2003,
  \mnras, 341, 669

\bibitem[{{Han} {et~al.}(2002){Han}, {Podsiadlowski}, {Maxted}, {Marsh}, \&
  {Ivanova}}]{Han2002}
{Han}, Z., {Podsiadlowski}, P., {Maxted}, P.~F.~L., {Marsh}, T.~R., \&
  {Ivanova}, N. 2002, \mnras, 336, 449

\bibitem[{{Hartman} \& {Bakos}(2016)}]{HartmanBakos2016}
{Hartman}, J.~D. \& {Bakos}, G.~{\'A}. 2016, Astronomy and Computing, 17, 1

\bibitem[{{Heber}(2016)}]{Heber2016}
{Heber}, U. 2016, \pasp, 128, 082001

\bibitem[{{Heber} {et~al.}(2013){Heber}, {Geier}, \& {Gaensicke}}]{Heber2013}
{Heber}, U., {Geier}, S., \& {Gaensicke}, B. 2013, in European Physical Journal
  Web of Conferences, Vol.~43, European Physical Journal Web of Conferences,
  04002

\bibitem[{{Henden} {et~al.}(2015){Henden}, {Levine}, {Terrell}, \&
  {Welch}}]{Henden2015}
{Henden}, A.~A., {Levine}, S., {Terrell}, D., \& {Welch}, D.~L. 2015, in
  American Astronomical Society Meeting Abstracts, Vol. 225, American
  Astronomical Society Meeting Abstracts \#225, 336.16

\bibitem[{{Howarth}(1987)}]{Howarth1987}
{Howarth}, I.~D. 1987, \mnras, 225, 33P

\bibitem[{{Hubeny} \& {Lanz}(1995)}]{Hubeny1995}
{Hubeny}, I. \& {Lanz}, T. 1995, \apj, 439, 875

\bibitem[{{Hubeny} \& {Lanz}(2017)}]{Hubeny2017}
{Hubeny}, I. \& {Lanz}, T. 2017, arXiv e-prints, arXiv:1706.01859

\bibitem[{{Iben}(1990)}]{Iben1990}
{Iben}, Icko, J. 1990, \apj, 353, 215

\bibitem[{{Iben} \& {Tutukov}(1986)}]{Iben1986}
{Iben}, Icko, J. \& {Tutukov}, A.~V. 1986, \apj, 311, 742

\bibitem[{{Igoshev} {et~al.}(2020){Igoshev}, {Perets}, \&
  {Michaely}}]{Igoshev2020}
{Igoshev}, A.~P., {Perets}, H.~B., \& {Michaely}, E. 2020, \mnras, 494, 1448

\bibitem[{{Jagelka} {et~al.}(2019){Jagelka}, {Mikul{\'a}{\v{s}}ek},
  {H{\"u}mmerich}, \& {Paunzen}}]{Jagelka2019}
{Jagelka}, M., {Mikul{\'a}{\v{s}}ek}, Z., {H{\"u}mmerich}, S., \& {Paunzen}, E.
  2019, \aap, 622, A199

\bibitem[{{Jordan} {et~al.}(2012){Jordan}, {Bagnulo}, {Werner}, \&
  {O'Toole}}]{Jordan2012}
{Jordan}, S., {Bagnulo}, S., {Werner}, K., \& {O'Toole}, S.~J. 2012, \aap, 542,
  A64

\bibitem[{{Jordan} {et~al.}(2005){Jordan}, {Werner}, \& {O'Toole}}]{Jordan2005}
{Jordan}, S., {Werner}, K., \& {O'Toole}, S.~J. 2005, \aap, 432, 273

\bibitem[{{Kepler} {et~al.}(2013){Kepler}, {Pelisoli}, {Jordan}, {Kleinman},
  {Koester}, {K{\"u}lebi}, {Pe{\c{c}}anha}, {Castanheira}, {Nitta}, {Costa},
  {Winget}, {Kanaan}, \& {Fraga}}]{Kepler2013}
{Kepler}, S.~O., {Pelisoli}, I., {Jordan}, S., {et~al.} 2013, \mnras, 429, 2934

\bibitem[{{K{\H{o}}v{\'a}ri} {et~al.}(2019){K{\H{o}}v{\'a}ri}, {Strassmeier},
  {Ol{\'a}h}, {Kriskovics}, {Vida}, {Carroll}, {Granzer}, {Ilyin}, {Jurcsik},
  {K{\H{o}}v{\'a}ri}, \& {Weber}}]{Kovari+2019}
{K{\H{o}}v{\'a}ri}, Z., {Strassmeier}, K.~G., {Ol{\'a}h}, K., {et~al.} 2019,
  \aap, 624, A83

\bibitem[{{Kupfer} {et~al.}(2015){Kupfer}, {Geier}, {Heber}, {{\O}stensen},
  {Barlow}, {Maxted}, {Heuser}, {Schaffenroth}, \& {G{\"a}nsicke}}]{Kupfer2015}
{Kupfer}, T., {Geier}, S., {Heber}, U., {et~al.} 2015, \aap, 576, A44

\bibitem[{{Kurucz}(1979)}]{Kurucz1979}
{Kurucz}, R.~L. 1979, \apjs, 40, 1

\bibitem[{{Lallement} {et~al.}(2019){Lallement}, {Babusiaux}, {Vergely},
  {Katz}, {Arenou}, {Valette}, {Hottier}, \& {Capitanio}}]{Lallement2019}
{Lallement}, R., {Babusiaux}, C., {Vergely}, J.~L., {et~al.} 2019, \aap, 625,
  A135

\bibitem[{{Landstreet}(2004)}]{Landstreet2004}
{Landstreet}, J.~D. 2004, in The A-Star Puzzle, ed. J.~{Zverko},
  J.~{Ziznovsky}, S.~J. {Adelman}, \& W.~W. {Weiss}, Vol. 224, 423--432

\bibitem[{{Landstreet} {et~al.}(2012){Landstreet}, {Bagnulo}, {Fossati},
  {Jordan}, \& {O'Toole}}]{Landstreet2012}
{Landstreet}, J.~D., {Bagnulo}, S., {Fossati}, L., {Jordan}, S., \& {O'Toole},
  S.~J. 2012, \aap, 541, A100

\bibitem[{{Landstreet} \& {Borra}(1978)}]{Landstreet+Borra1978}
{Landstreet}, J.~D. \& {Borra}, E.~F. 1978, \apjl, 224, L5

\bibitem[{{Latour} {et~al.}(2015){Latour}, {Fontaine}, {Green}, \&
  {Brassard}}]{Latour2015}
{Latour}, M., {Fontaine}, G., {Green}, E.~M., \& {Brassard}, P. 2015, \aap,
  579, A39

\bibitem[{{Leone} {et~al.}(2014){Leone}, {Corradi}, {Mart{\'\i}nez
  Gonz{\'a}lez}, {Asensio Ramos}, \& {Manso Sainz}}]{Leone2014}
{Leone}, F., {Corradi}, R.~L.~M., {Mart{\'\i}nez Gonz{\'a}lez}, M.~J., {Asensio
  Ramos}, A., \& {Manso Sainz}, R. 2014, \aap, 563, A43

\bibitem[{{Li} {et~al.}(1998){Li}, {Ferrario}, \& {Wickramasinghe}}]{Li+1998}
{Li}, J., {Ferrario}, L., \& {Wickramasinghe}, D. 1998, \apjl, 503, L151

\bibitem[{{Lightkurve Collaboration} {et~al.}(2018){Lightkurve Collaboration},
  {Cardoso}, {Hedges}, {Gully-Santiago}, {Saunders}, {Cody}, {Barclay}, {Hall},
  {Sagear}, {Turtelboom}, {Zhang}, {Tzanidakis}, {Mighell}, {Coughlin}, {Bell},
  {Berta-Thompson}, {Williams}, {Dotson}, \& {Barentsen}}]{lightkurve}
{Lightkurve Collaboration}, {Cardoso}, J.~V.~d.~M., {Hedges}, C., {et~al.}
  2018, {Lightkurve: Kepler and TESS time series analysis in Python},
  Astrophysics Source Code Library

\bibitem[{{Lillo-Box} {et~al.}(2014){Lillo-Box}, {Barrado}, {Moya},
  {Montesinos}, {Montalb{\'a}n}, {Bayo}, {Barbieri}, {R{\'e}gulo}, {Mancini},
  {Bouy}, \& {Henning}}]{Lillobox2014}
{Lillo-Box}, J., {Barrado}, D., {Moya}, A., {et~al.} 2014, \aap, 562, A109

\bibitem[{{Lindegren} {et~al.}(2020{\natexlab{a}}){Lindegren}, {Bastian},
  {Biermann}, {Bombrun}, {de Torres}, {Gerlach}, {Geyer}, {Hern{\'a}ndez},
  {Hilger}, {Hobbs}, {Klioner}, {Lammers}, {McMillan}, {Ramos-Lerate},
  {Steidelm{\"u}ller}, {Stephenson}, \& {van Leeuwen}}]{Lindegren2020b}
{Lindegren}, L., {Bastian}, U., {Biermann}, M., {et~al.} 2020{\natexlab{a}},
  arXiv e-prints, arXiv:2012.01742

\bibitem[{{Lindegren} {et~al.}(2020{\natexlab{b}}){Lindegren}, {Klioner},
  {Hern{\'a}ndez}, {Bombrun}, {Ramos-Lerate}, {Steidelm{\"u}ller}, {Bastian},
  {Biermann}, {de Torres}, {Gerlach}, {Geyer}, {Hilger}, {Hobbs}, {Lammers},
  {McMillan}, {Stephenson}, {Casta{\~n}eda}, {Davidson}, {Fabricius},
  {Gracia-Abril}, {Portell}, {Rowell}, {Teyssier}, {Torra}, {Bartolom{\'e}},
  {Clotet}, {Garralda}, {Gonz{\'a}lez-Vidal}, {Torra}, {Abbas}, {Altmann},
  {Anglada Varela}, {Balaguer-N{\'u}{\~n}ez}, {Balog}, {Barache}, {Becciani},
  {Bernet}, {Bertone}, {Bianchi}, {Bouquillon}, {Brown}, {Bucciarelli},
  {Busonero}, {Butkevich}, {Buzzi}, {Cancelliere}, {Carlucci}, {Charlot},
  {Cioni}, {Crosta}, {Crowley}, {del Peloso}, {del Pozo}, {Drimmel}, {Esquej},
  {Fienga}, {Fraile}, {Gai}, {Garcia-Reinaldos}, {Guerra}, {Hambly}, {Hauser},
  {Jan{\ss}en}, {Jordan}, {Kostrzewa-Rutkowska}, {Lattanzi}, {Liao}, {Licata},
  {Lister}, {L{\"o}ffler}, {Marchant}, {Masip}, {Mignard}, {Mints}, {Molina},
  {Mora}, {Morbidelli}, {Murphy}, {Pagani}, {Panuzzo}, {Pe{\~n}alosa Esteller},
  {Poggio}, {Re Fiorentin}, {Riva}, {Sagrist{\`a} Sell{\'e}s}, {Sanchez
  Gimenez}, {Sarasso}, {Sciacca}, {Siddiqui}, {Smart}, {Souami}, {Spagna},
  {Steele}, {Taris}, {Utrilla}, {van Reeven}, \& {Vecchiato}}]{Lindegren2020}
{Lindegren}, L., {Klioner}, S.~A., {Hern{\'a}ndez}, J., {et~al.}
  2020{\natexlab{b}}, arXiv e-prints, arXiv:2012.03380

\bibitem[{{Luo} {et~al.}(2020){Luo}, {N{\'e}meth}, \& {Li}}]{Luo2020}
{Luo}, Y., {N{\'e}meth}, P., \& {Li}, Q. 2020, \apj, 898, 64

\bibitem[{{Mardling} \& {Aarseth}(1999)}]{Mardling1999}
{Mardling}, R. \& {Aarseth}, S. 1999, in NATO Advanced Study Institute (ASI)
  Series C, Vol. 522, The Dynamics of Small Bodies in the Solar System, A Major
  Key to Solar System Studies, ed. B.~A. {Steves} \& A.~E. {Roy}, 385

\bibitem[{{Mathys} {et~al.}(2012){Mathys}, {Hubrig}, {Mason}, {Michaud},
  {Sch{\"o}ller}, \& {Wesemael}}]{Mathys2012}
{Mathys}, G., {Hubrig}, S., {Mason}, E., {et~al.} 2012, Astronomische
  Nachrichten, 333, 30

\bibitem[{{Momany} {et~al.}(2020){Momany}, {Zaggia}, {Montalto}, {Jones},
  {Boffin}, {Cassisi}, {Moni Bidin}, {Gullieuszik}, {Saviane}, {Monaco},
  {Mason}, {Girardi}, {D'Orazi}, {Piotto}, {Milone}, {Lala}, {Stetson}, \&
  {Beletsky}}]{Momany2020}
{Momany}, Y., {Zaggia}, S., {Montalto}, M., {et~al.} 2020, Nature Astronomy, 4,
  1092

\bibitem[{{Neiner} {et~al.}(2015){Neiner}, {Mathis}, {Alecian}, {Emeriau},
  {Grunhut}, {BinaMIcS}, \& {MiMeS Collaborations}}]{Neiner2015}
{Neiner}, C., {Mathis}, S., {Alecian}, E., {et~al.} 2015, in Polarimetry, ed.
  K.~N. {Nagendra}, S.~{Bagnulo}, R.~{Centeno}, \& M.~{Jes{\'u}s Mart{\'\i}nez
  Gonz{\'a}lez}, Vol. 305, 61--66

\bibitem[{{O'Toole} {et~al.}(2005){O'Toole}, {Jordan}, {Friedrich}, \&
  {Heber}}]{OToole2005}
{O'Toole}, S.~J., {Jordan}, S., {Friedrich}, S., \& {Heber}, U. 2005, \aap,
  437, 227

\bibitem[{{Pauli} {et~al.}(2006){Pauli}, {Napiwotzki}, {Heber}, {Altmann}, \&
  {Odenkirchen}}]{Pauli2006}
{Pauli}, E.~M., {Napiwotzki}, R., {Heber}, U., {Altmann}, M., \& {Odenkirchen},
  M. 2006, \aap, 447, 173

\bibitem[{{Pelisoli} {et~al.}(2020){Pelisoli}, {Vos}, {Geier}, {Schaffenroth},
  \& {Baran}}]{Pelisoli2020}
{Pelisoli}, I., {Vos}, J., {Geier}, S., {Schaffenroth}, V., \& {Baran}, A.~S.
  2020, arXiv e-prints, arXiv:2008.07522

\bibitem[{{Petit} {et~al.}(2013){Petit}, {Owocki}, {Wade}, {Cohen},
  {Sundqvist}, {Gagn{\'e}}, {Ma{\'{\i}}z Apell{\'a}niz}, {Oksala}, {Bohlender},
  {Rivinius}, {Henrichs}, {Alecian}, {Townsend}, {ud-Doula}, \& {MiMeS
  Collaboration}}]{Petit+2013}
{Petit}, V., {Owocki}, S.~P., {Wade}, G.~A., {et~al.} 2013, \mnras, 429, 398

\bibitem[{{Potter} \& {Tout}(2010)}]{Potter2010}
{Potter}, A.~T. \& {Tout}, C.~A. 2010, \mnras, 402, 1072

\bibitem[{{Press} {et~al.}(1992){Press}, {Teukolsky}, {Vetterling}, \&
  {Flannery}}]{Press1992}
{Press}, W.~H., {Teukolsky}, S.~A., {Vetterling}, W.~T., \& {Flannery}, B.~P.
  1992, {Numerical recipes in C. The art of scientific computing}

\bibitem[{{Prince} {et~al.}(2019){Prince}, {Burdge}, {Bellm}, {Coughlin},
  {Kaplan}, {Kupfer}, \& {van Roestel}}]{Prince2019}
{Prince}, T., {Burdge}, K., {Bellm}, E., {et~al.} 2019, in American
  Astronomical Society Meeting Abstracts, Vol. 233, American Astronomical
  Society Meeting Abstracts \#233, 418.05

\bibitem[{{Rauch} {et~al.}(2010){Rauch}, {Werner}, \& {Kruk}}]{Rauch2010}
{Rauch}, T., {Werner}, K., \& {Kruk}, J.~W. 2010, \apss, 329, 133

\bibitem[{{Reding} {et~al.}(2020){Reding}, {Hermes}, {Vanderbosch}, {Dennihy},
  {Kaiser}, {Mace}, {Dunlap}, \& {Clemens}}]{Reding+2020}
{Reding}, J.~S., {Hermes}, J.~J., {Vanderbosch}, Z., {et~al.} 2020, \apj, 894,
  19

\bibitem[{{Reindl} {et~al.}(2014){Reindl}, {Rauch}, {Werner}, {Kruk}, \&
  {Todt}}]{Reindl2014}
{Reindl}, N., {Rauch}, T., {Werner}, K., {Kruk}, J.~W., \& {Todt}, H. 2014,
  \aap, 566, A116

\bibitem[{{Ricker} {et~al.}(2015){Ricker}, {Winn}, {Vanderspek}, {Latham},
  {Bakos}, {Bean}, {Berta-Thompson}, {Brown}, {Buchhave}, {Butler}, {Butler},
  {Chaplin}, {Charbonneau}, {Christensen-Dalsgaard}, {Clampin}, {Deming},
  {Doty}, {De Lee}, {Dressing}, {Dunham}, {Endl}, {Fressin}, {Ge}, {Henning},
  {Holman}, {Howard}, {Ida}, {Jenkins}, {Jernigan}, {Johnson}, {Kaltenegger},
  {Kawai}, {Kjeldsen}, {Laughlin}, {Levine}, {Lin}, {Lissauer}, {MacQueen},
  {Marcy}, {McCullough}, {Morton}, {Narita}, {Paegert}, {Palle}, {Pepe},
  {Pepper}, {Quirrenbach}, {Rinehart}, {Sasselov}, {Sato}, {Seager},
  {Sozzetti}, {Stassun}, {Sullivan}, {Szentgyorgyi}, {Torres}, {Udry}, \&
  {Villasenor}}]{ricker2015}
{Ricker}, G.~R., {Winn}, J.~N., {Vanderspek}, R., {et~al.} 2015, Journal of
  Astronomical Telescopes, Instruments, and Systems, 1, 014003

\bibitem[{{Riello} {et~al.}(2020){Riello}, {De Angeli}, {Evans}, {Montegriffo},
  \& et~al.}]{Riello2020}
{Riello}, M., {De Angeli}, F., {Evans}, D., {Montegriffo}, P., \& et~al. 2020,
  Submitted to \aap

\bibitem[{{Sabin} {et~al.}(2015){Sabin}, {Hull}, {Plambeck}, {Zijlstra},
  {V{\'a}zquez}, {Navarro}, \& {Guill{\'e}n}}]{Sabin2015}
{Sabin}, L., {Hull}, C.~L.~H., {Plambeck}, R.~L., {et~al.} 2015, \mnras, 449,
  2368

\bibitem[{{Saio} \& {Jeffery}(2000)}]{Saio2000}
{Saio}, H. \& {Jeffery}, C.~S. 2000, \mnras, 313, 671

\bibitem[{{Schimeczek} \& {Wunner}(2014)}]{Schimeczek2014}
{Schimeczek}, C. \& {Wunner}, G. 2014, \apjs, 212, 26

\bibitem[{{Schlafly} {et~al.}(2019){Schlafly}, {Meisner}, \&
  {Green}}]{Schlafly2019}
{Schlafly}, E.~F., {Meisner}, A.~M., \& {Green}, G.~M. 2019, \apjs, 240, 30

\bibitem[{{Schneider} {et~al.}(2019){Schneider}, {Ohlmann}, {Podsiadlowski},
  {R{\"o}pke}, {Balbus}, {Pakmor}, \& {Springel}}]{Schneider2019}
{Schneider}, F. R.~N., {Ohlmann}, S.~T., {Podsiadlowski}, P., {et~al.} 2019,
  \nat, 574, 211

\bibitem[{{Schwab}(2018)}]{Schwab2018}
{Schwab}, J. 2018, \mnras, 476, 5303

\bibitem[{{Shultz} {et~al.}(2020){Shultz}, {Owocki}, {Rivinius}, {Wade},
  {Neiner}, {Alecian}, {Kochukhov}, {Bohlender}, {ud-Doula}, {Landstreet},
  {Sikora}, {David-Uraz}, {Petit}, {Cerraho{\u{g}}lu}, {Fine}, {Henson}, {MiMeS
  Collaboration}, \& {BinaMIcS Collaboration}}]{Shultz+2020}
{Shultz}, M.~E., {Owocki}, S., {Rivinius}, T., {et~al.} 2020, \mnras, 499, 5379

\bibitem[{{Skrutskie} {et~al.}(2006){Skrutskie}, {Cutri}, {Stiening},
  {Weinberg}, {Schneider}, {Carpenter}, {Beichman}, {Capps}, {Chester},
  {Elias}, {Huchra}, {Liebert}, {Lonsdale}, {Monet}, {Price}, {Seitzer},
  {Jarrett}, {Kirkpatrick}, {Gizis}, {Howard}, {Evans}, {Fowler}, {Fullmer},
  {Hurt}, {Light}, {Kopan}, {Marsh}, {McCallon}, {Tam}, {Van Dyk}, \&
  {Wheelock}}]{Skrutskie2006}
{Skrutskie}, M.~F., {Cutri}, R.~M., {Stiening}, R., {et~al.} 2006, \aj, 131,
  1163

\bibitem[{{Ter Braak}(2006)}]{TerBraa+Cajok2006}
{Ter Braak}, C. J.~F. 2006, Statistics and Computing, 16, 239

\bibitem[{{Toonen} {et~al.}(2016){Toonen}, {Hamers}, \& {Portegies
  Zwart}}]{Toonen2016}
{Toonen}, S., {Hamers}, A., \& {Portegies Zwart}, S. 2016, Computational
  Astrophysics and Cosmology, 3, 6

\bibitem[{{Toonen} {et~al.}(2020){Toonen}, {Portegies Zwart}, {Hamers}, \&
  {Bandopadhyay}}]{Toonen2020}
{Toonen}, S., {Portegies Zwart}, S., {Hamers}, A.~S., \& {Bandopadhyay}, D.
  2020, \aap, 640, A16

\bibitem[{{Tout} {et~al.}(2008){Tout}, {Wickramasinghe}, {Liebert}, {Ferrario},
  \& {Pringle}}]{Tout2008}
{Tout}, C.~A., {Wickramasinghe}, D.~T., {Liebert}, J., {Ferrario}, L., \&
  {Pringle}, J.~E. 2008, \mnras, 387, 897

\bibitem[{{Townsend} \& {Owocki}(2005)}]{Townsend+Owocki2005}
{Townsend}, R.~H.~D. \& {Owocki}, S.~P. 2005, \mnras, 357, 251

\bibitem[{{Townsend} {et~al.}(2007){Townsend}, {Owocki}, \&
  {Ud-Doula}}]{Townsend+2007}
{Townsend}, R.~H.~D., {Owocki}, S.~P., \& {Ud-Doula}, A. 2007, \mnras, 382, 139

\bibitem[{{Ud-Doula} \& {Owocki}(2002)}]{Ud-Doula+2002}
{Ud-Doula}, A. \& {Owocki}, S.~P. 2002, \apj, 576, 413

\bibitem[{{Ud-Doula} {et~al.}(2008){Ud-Doula}, {Owocki}, \&
  {Townsend}}]{Ud-Doula+2008}
{Ud-Doula}, A., {Owocki}, S.~P., \& {Townsend}, R. H.~D. 2008, \mnras, 385, 97

\bibitem[{{Vida} {et~al.}(2015){Vida}, {Korhonen}, {Ilyin}, {Ol{\'a}h},
  {Andersen}, \& {Hackman}}]{Vida+2015}
{Vida}, K., {Korhonen}, H., {Ilyin}, I.~V., {et~al.} 2015, \aap, 580, A64

\bibitem[{{Vink} \& {Cassisi}(2002)}]{VinkCassisi2002}
{Vink}, J.~S. \& {Cassisi}, S. 2002, \aap, 392, 553

\bibitem[{{Vos} {et~al.}(2020){Vos}, {Bobrick}, \&
  {Vu{\v{c}}kovi{\'c}}}]{Vos2020}
{Vos}, J., {Bobrick}, A., \& {Vu{\v{c}}kovi{\'c}}, M. 2020, \aap, 641, A163

\bibitem[{{Vos} {et~al.}(2012){Vos}, {{\O}stensen}, {Degroote}, {De Smedt},
  {Green}, {Heber}, {Van Winckel}, {Acke}, {Bloemen}, {De Cat}, {Exter},
  {Lampens}, {Lombaert}, {Masseron}, {Menu}, {Neyskens}, {Raskin}, {Ringat},
  {Rauch}, {Smolders}, \& {Tkachenko}}]{Vos2012}
{Vos}, J., {{\O}stensen}, R.~H., {Degroote}, P., {et~al.} 2012, \aap, 548, A6

\bibitem[{{Vos} {et~al.}(2013){Vos}, {{\O}stensen}, {N{\'e}meth}, {Green},
  {Heber}, \& {Van Winckel}}]{Vos2013}
{Vos}, J., {{\O}stensen}, R.~H., {N{\'e}meth}, P., {et~al.} 2013, \aap, 559,
  A54

\bibitem[{{Vos} {et~al.}(2017){Vos}, {{\O}stensen}, {Vu{\v{c}}kovi{\'c}}, \&
  {Van Winckel}}]{Vos2017}
{Vos}, J., {{\O}stensen}, R.~H., {Vu{\v{c}}kovi{\'c}}, M., \& {Van Winckel}, H.
  2017, \aap, 605, A109

\bibitem[{{Webbink}(1984)}]{Webbink1984}
{Webbink}, R.~F. 1984, \apj, 277, 355

\bibitem[{{Werner} {et~al.}(2003){Werner}, {Deetjen}, {Dreizler}, {Nagel},
  {Rauch}, \& {Schuh}}]{Werner2003}
{Werner}, K., {Deetjen}, J.~L., {Dreizler}, S., {et~al.} 2003, in ASPCS, Vol.
  288, Stellar Atmosphere Modeling, ed. I.~{Hubeny}, D.~{Mihalas}, \&
  K.~{Werner}, 31

\bibitem[{{Werner} {et~al.}(2020){Werner}, {Reindl}, {L{\"o}bling}, {Pelisoli},
  {Schaffenroth}, {Rebassa-Mansergas}, {Irawati}, \& {Ren}}]{Werner+2020}
{Werner}, K., {Reindl}, N., {L{\"o}bling}, L., {et~al.} 2020, \aap, 642, A228

\bibitem[{{Wickramasinghe} {et~al.}(2010){Wickramasinghe}, {Farihi}, {Tout},
  {Ferrario}, \& {Stancliffe}}]{Wickramasinghe+2010}
{Wickramasinghe}, D.~T., {Farihi}, J., {Tout}, C.~A., {Ferrario}, L., \&
  {Stancliffe}, R.~J. 2010, \mnras, 404, 1984

\bibitem[{{Wilson}(1963)}]{Wilson1963}
{Wilson}, O.~C. 1963, \apj, 138, 832

\bibitem[{{Wilson}(1968)}]{Wilson1968}
{Wilson}, O.~C. 1968, \apj, 153, 221

\bibitem[{{Wolf} {et~al.}(2018){Wolf}, {Onken}, {Luvaul}, {Schmidt}, {Bessell},
  {Chang}, {Da Costa}, {Mackey}, {Martin-Jones}, {Murphy}, {Preston}, {Scalzo},
  {Shao}, {Smillie}, {Tisserand}, {White}, \& {Yuan}}]{Wolf2018}
{Wolf}, C., {Onken}, C.~A., {Luvaul}, L.~C., {et~al.} 2018, \pasa, 35, e010

\bibitem[{{Zechmeister} \& {K{\"u}rster}(2009)}]{ZechmeisterKuerster2009}
{Zechmeister}, M. \& {K{\"u}rster}, M. 2009, \aap, 496, 577

\bibitem[{{Zhang} {et~al.}(2017){Zhang}, {Hall}, {Jeffery}, \&
  {Bi}}]{Zhang2017}
{Zhang}, X., {Hall}, P.~D., {Jeffery}, C.~S., \& {Bi}, S. 2017, \apj, 835, 242

\bibitem[{{Zhang} \& {Jeffery}(2012)}]{Zhang2012}
{Zhang}, X. \& {Jeffery}, C.~S. 2012, \mnras, 419, 452

\end{thebibliography}

\begin{appendix}

\section{XSHOOTER spectra}

\begin{figure*}
    \includegraphics[width=\linewidth]{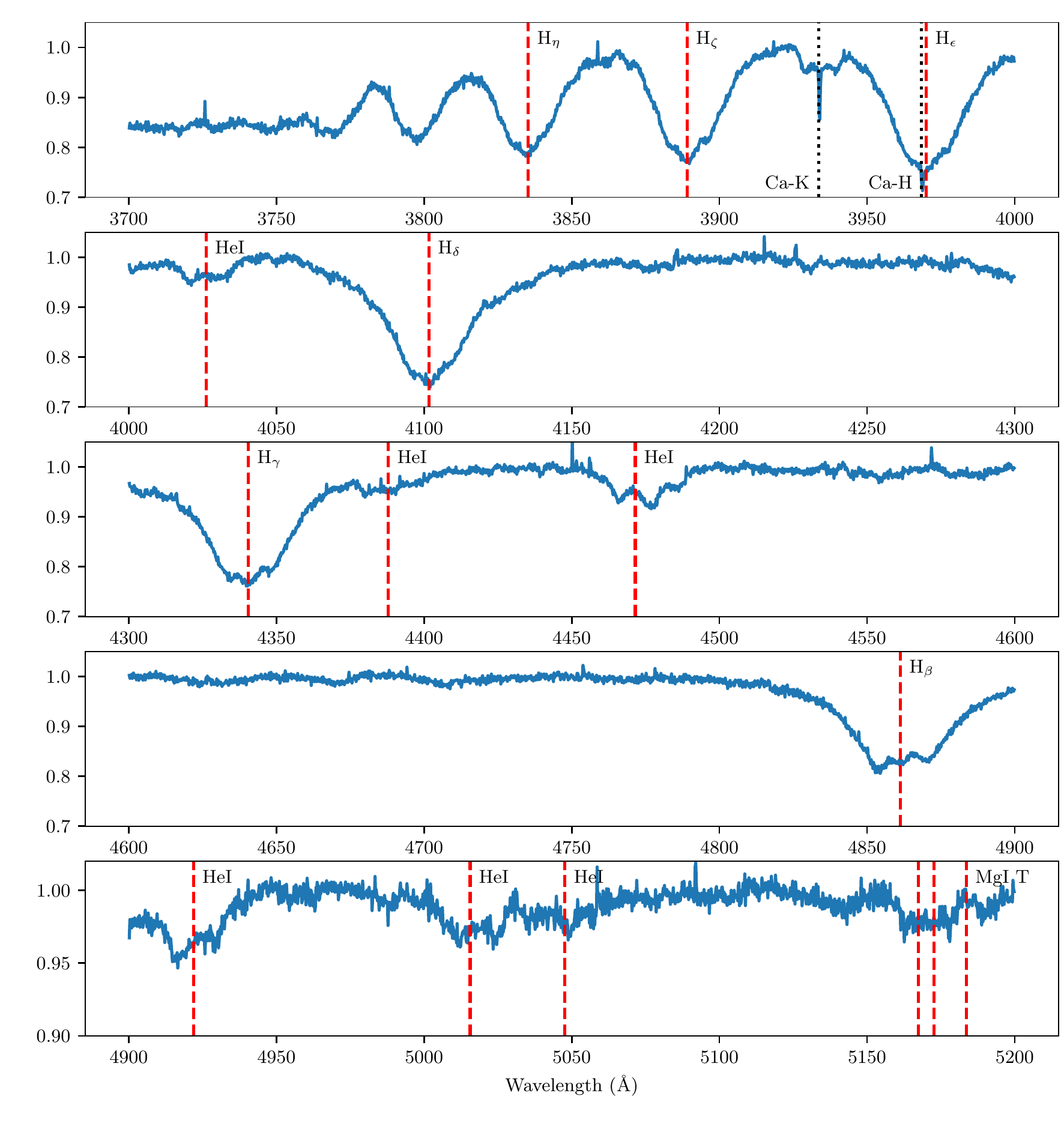}
    \caption{The merged X-SHOOTER spectrum of J22564-5910, showing the UVB arm. Spectral features of interest of the star are indicated by vertical red dashed lines. The sharp interstellar Ca lines are marked in black. The location of the \ion{Mg}{i} triplet is marked, but the quality of the spectra is not sufficient to confirm its detection.}\label{fig:xshooter_merged_normalized_UVB}
\end{figure*}

\begin{figure*}
    \includegraphics[width=\linewidth]{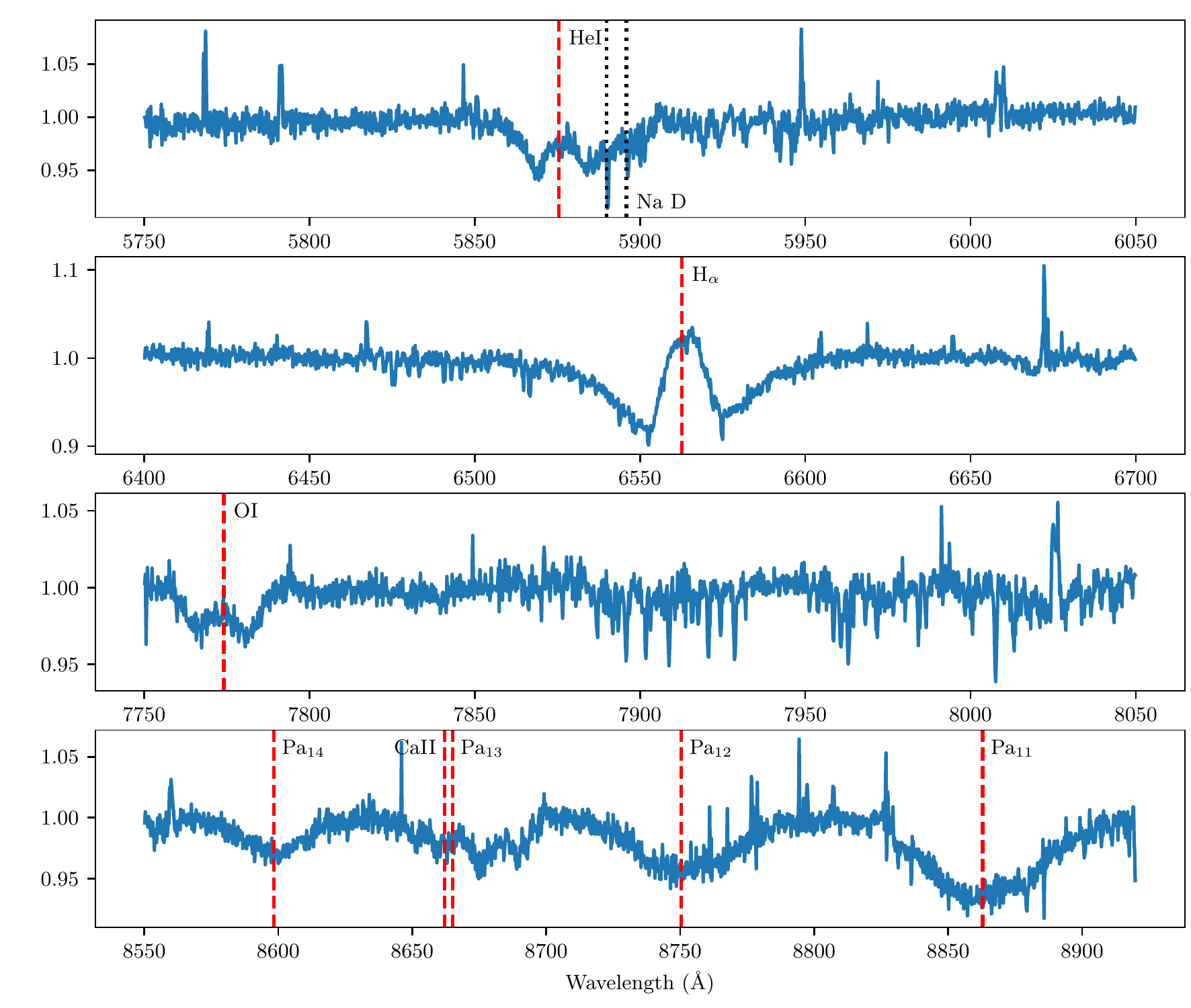}
    \caption{The merged X-SHOOTER spectrum of J22564-5910, showing the VIS arm. Spectral features of interest of the star are indicated by vertical red dashed lines. The interstellar sharp sodium doublet is marked in black. The Pachen H lines are indicated by `Pa$_{\rm nn}$'. The location of the $\lambda 8662$ line of the Ca IR triple is shown in the bottom row. The other two lines of the Ca IR triplet are not visible in the spectrum.}\label{fig:xshooter_merged_normalized_VIS}
\end{figure*}

\section{Nearby companion}\label{ap:resolved-companion}

\begin{table}[t]
    \centering
    \caption{The parallax and proper motion of \J\ and its nearby companion, obtained from Gaia EDR3.}\label{tb:gaia_dr3}
    \begin{tabular}{lllll}
    \hline\hline
    \noalign{\smallskip}
     & Parallax & PM RA & PM DEC & Gaia-G\\
     & (mas) & (mas/yr) & (mas/yr) & (mag)\\
    \hline
    \noalign{\smallskip}
    1 & 1.58 $\pm$ 0.02 & 13.77 $\pm$ 0.02 & -24.43 $\pm$ 0.02 & 14.24 \\
    2 & 1.52 $\pm$ 0.13 & -5.81 $\pm$ 0.11 & -23.90 $\pm$ 0.12 & 18.40 \\
    \hline
    \end{tabular}
    \tablefoot{Star 1 = \J\ = Gaia EDR3 6491685395361112832, star 2 = Gaia EDR3 6491685391064878080}.
\end{table}

Images show a relatively nearby star (Gaia EDR3 6491685391064878080) at a separation of $\sim$~$17.2$ arcsec. The parallax and proper motion of both stars are given in Table\,\ref{tb:gaia_dr3}. The parallax of the nearby star is nearly identical to that of \J; $1.58 \pm 0.02$ and $1.52 \pm 0.13$ respectively for \J\ and the companion. 
At a distance of $635$ pc, the separation between both systems is roughly $8300$ AU, which is fairly common for sdBs with wide astrometric companions \citep{Igoshev2020}. The SED of the companion star hints at a very cool small star (T$_{\rm eff} < 3500 K$, R $\sim 0.5$ R$_{\odot}$). This companion star is too far away and too faint (Gaia G = $18.4$ mag, 2MASS J = $15.7$ mag) to influence the observations of \J.

The difference in the proper motion between \J\ and the companion corresponds to a physical velocity of 58.7\,km/s.  At the same time, the escape velocity for a 0.5+0.5 M$_{\odot}$ binary with a separation of 8300\,AU is equal to $\sim$0.33\,km/s. This strongly suggests that the objects are unbound from each other. When tracing their galactic orbits backwards, they cross but not at the same time. It is possible to interpret this as a scenario where the system started out as a hierarchical triple. The inner binary merges to form the sdB, and due to the merger process, the outer companion gets ejected. In this case, the distance and velocity difference can be used to estimate the time passed since the merger. Assuming the trajectories of the two stars share the same origin, the travel time would be about $670\,{\rm yr}$. \J\ would then be observed very early after the merger. However, since the orbits do not place both components at exactly the same position at the same time, it is possible that this is just a chance encounter.

The presence of a nearby optical companion can indicate a possible triple origin of the system, \citep[e.g.][]{Toonen2016}. In this case, the MS companion would be ejected during a dynamical triple interaction phase between the MS companion and two He-WDs. \J\ would then form as a remnant from a merger of these two He-WDs. In this case, the system would have to be only $\sim700\,{\rm yr}$ old, which might explain the presence of disk-like CM or the strong magnetic field of the sdB, which would be then driven by trace accretion. Such dynamical triple interaction could have been triggered by mass transfer between an RGB star and a He-WD companion. Mass loss from the more massive red giant would widen the inner orbit and, if the tertiary companion were sufficiently close, drive the system made of the tertiary MS star, the He-WD accretor and the core of the red giant towards a chaotic dynamical triple interaction phase \citep{Mardling1999, Toonen2020}. At the end of this phase, the He-WDs would merge, leaving an unbound MS companion. This scenario requires that some red giant material lost during the mass transfer phase would remain {\it between} \J\ and its companion, not bound to either star in particular, which may possibly be detectable with ALMA. Similarly, this scenario requires that in radial velocity, the MS companion is moving away from \J. 

\end{appendix}

\end{document}